\documentclass[11pt]{article}
\usepackage{amsfonts,amssymb,amsmath}
\usepackage{cite}
\usepackage{graphicx}
\newcounter{tempeq}
\usepackage{amsmath,amssymb,color,graphicx,multirow,young,float,rotfloat }
\usepackage[flushleft]{threeparttable}
\usepackage[vcentermath]{youngtab}
\usepackage{hyperref}
\usepackage{caption}
\usepackage{xcolor}
\definecolor{darkred}{rgb}{0.65,0.15,0}
\hypersetup{pdfborder={0 0 0},colorlinks=true,urlcolor=darkred,citecolor=blue,linkcolor=darkred,linktocpage=true}

\usepackage{cite,amsfonts,amsthm,mathrsfs}
\usepackage{lscape}
\usepackage[T1]{fontenc}
\usepackage{enumerate}
\hypersetup{
	colorlinks   =  true,
	citecolor    = red,
	urlcolor	=magenta,
}

\def\4diml{four-dimensional}

\def\-1{^{-1}}

\makeatletter
\renewcommand{\theequation}{\thesection.\arabic{equation}}
\@addtoreset{equation}{section}
\makeatother

\begin{document}
\title{\bf 4+1 D homogeneous anisotropic string cosmological models with dilaton and anti-symmetric matter}
\author{R. Aslefatollahi{\thanks{e-mail:
fatollahie@azaruniv.ac.ir}}\,\,\,B. Mojaveri{\thanks{e-mail:
bmojaveri@azaruniv.ac.ir}}\,\,\, A. Rezaei-Aghdam{\thanks{
coresponding author, e-mail: rezaei-a@azaruniv.ac.ir}}\\ {\small
 {\em Department of Physics, Azarbaijan Shahid Madani University,
53714-161, Tabriz, Iran.} }}\maketitle
\begin{abstract}
We proceed to investigate exact solutions of homogeneous anisotropic
string cosmological models characterized by five-dimensional
space-time metrics (with real four-dimensional isometry Lie groups),
non-vanishing dilaton and anti-symmetric B field.
\end{abstract}
\section{\bf Introduction} It seems that the low energy string effective action\cite{Frad,Ca} is a great model for study of the dynamics of the very early universe below the string
scale\cite{A,BV,V,GV1}. There are several reasons that show the dynamic of early
universe below the plank or string scale has been influenced by
 the presence of spatial anisotropies (for a good review on string cosmology see ref\cite{GV2,L} and the book \cite{G})\footnote{Also for a good bibliography until 2014 see the Maurizio Gasperini's home page https://home.ba.infn.it/~gasperin/.} . It has generally claimed that isotropy of
universe is deduced from the observations and measurements of cosmic
microwave background,which is the accurate ones in cosmology.
However these results are valid for the times that universe became
transparent. Statistical fluctuation in Friedmann-Robertson-Walker
model does not confirm with the observed state of universe. This
problem could be resolved by string theory by effective limiting
space-time curvature. The spatially homogeneous anisotropic Bianchi
string cosmologies \cite{Gas,Ba,Ba1,Ba2,Barrow1} have been studied previously (for the inhomogeneous case see \cite{Barrow2}). Also the effect of duality on string cosmological backgrounds has been explored in several works \cite{GG}. Bianchi string cosmologies admit a three-dimensional real Lie group of isometries acting
simply-transitively on the three-dimensional space-like orbits
\cite{Ba}. A study on 2+1 dimensional homogeneous string cosmology with two-dimensional Lie algebras was performed in \cite{NR}. Additionally, a generalization to models with four-dimensional real Lie groups was also performed \cite{Mojaveri}. In the our previous work \cite{Mojaveri}, we obtained exact solutions for string cosmological models characterized by a five-dimensional metric (with four-dimensional real Lie groups as isometry groups), along with a space-independent dilaton field and vanishing torsion. In this paper, we will obtain exact solutions for non-vanishing torsion with different cases of their space components.
\newline The paper is organized as follows. In Section 2, we review homogeneous anisotropic string cosmology in $d+1$ dimensions \cite{Gas}. Furthermore, we obtain the main equations in $4+1$ dimensions at the end of this section and introduce four families of space-dependent anti-symmetric B-fields.
 In Section 3, we obtain the potential $V_i$ related to models with real four-dimensional Lie groups and attempt to find exact solutions for the equations of motion of homogeneous anisotropic string cosmological models for each family of B-fields. The results are summarized in Tables 1-4. In Section 4, a cosmological model with the Lie group $VII_0\bigoplus R$ is investigated as a model with a perfect fluid (dark energy). We demonstrate that the model can be reduced to a $3+1$-dimensional cosmological model, where the extra dimension is compactified.
The four-dimensional real Lie algebras \cite{Pat} and their related left invariants one forms \cite{Mojaveri} are provided in the Appendix A. Some corrections about the models of section 5 of ref \cite{Mojaveri} are given in Appendix B.
\section{\bf Review of homogeneous anisotropic string cosmology}
To present the notations, let us consider the low-energy string effective action on a $(d+1)$-dimensional target space\cite{Frad,Ca}
\renewcommand\theequation{\arabic{tempeq}\alph{equation}}
\setcounter{equation}{-1} \addtocounter{tempeq}{1}
\begin{equation}
S_{eff}=\int d^{d+1}x \sqrt{-g}e^{\phi}(R-\frac{1}{12}
H_{\mu\nu\rho}H^{\mu\nu\rho}+\partial_{\mu}\phi\partial^{\mu}\phi-\Lambda)
\label{b4'}.
\end{equation}
The equations of motion of the action can be expressed as \cite{Ca}
\renewcommand\theequation{\arabic{tempeq}\alph{equation}}
\setcounter{equation}{0} \addtocounter{tempeq}{1}
\begin{eqnarray}
R_{\mu\nu}-\frac{1}{4}H_{\mu\nu}^2-\nabla_\mu\nabla_\nu\phi&=&0,
\label{b1}\\
\nabla^{\mu}(e^\phi H_{\mu\nu\lambda})&=&0, \label{b2} \\
-R+\frac{1}{12}H^2+2\nabla^2
\phi+(\partial_\mu\phi)^2+\Lambda&=&0.\label{b3}
\end{eqnarray}
Here, $\phi$ represents the dilaton field, $\Lambda$ stands for the cosmological constant, $R_{\mu\nu}$ denotes the Ricci curvature, $R$ is the Ricci scalar, and the terms $H_{\mu\nu}^2=H_{\mu\kappa\lambda}
{H_{\nu}}^{\kappa\lambda} \, ,
H^2=H_{\mu\nu\lambda}H^{\mu\nu\lambda}$ involve the totally
anti-symmetric field strength $H_{\mu\nu\lambda}$ defined from
$B_{\mu\nu}$ field, as
\renewcommand\theequation{\arabic{tempeq}\alph{equation}}
\setcounter{equation}{-1} \addtocounter{tempeq}{1}
\begin{eqnarray}
H_{\mu\nu\rho}=\partial_\mu B_{\nu\rho}+\partial_\nu
B_{\rho\mu}+\partial_\rho B_{\mu\nu}.
\end{eqnarray}
 The equations $(2a)-(2c)$ represent the Einstein field equations coupled with matter fields $\phi$ and $H$ (see, for instance, \cite{Gas, Ba}). Assuming spatial homogeneity, the $d$-dimensional spatial submanifold remains invariant under the action of a $d$-dimensional isometry group $G$. As a result, the metric of the space-time $G_{\mu\nu}$ can be factorized in the following form \cite{Gas,La}:
\renewcommand\theequation{\arabic{tempeq}\alph{equation}}
\setcounter{equation}{-1} \addtocounter{tempeq}{1}
\begin{eqnarray}
ds^2=G_{\mu\nu}dx^{\mu}dx^{\nu}=-dt^2+{e_\mu}^i(x)g_{ij}(t){e_\nu}^j(x)dx^\mu dx^\nu,
\end{eqnarray}
Here, $\mu, \nu=0,1,...,d$ are space-time indices in the spatial submanifold, and $i,j=1,...,d$ are indices of the isometry Lie group generators. The vielbeins ${e_\mu}^i(x)$ depend solely on the spatial coordinates \cite{La}. In contrast we assume that the metric $g_{ij}$ on the Lie algebra $\mathcal{G}$ of $G$ is solely a function of time $t$. To maintain homogeneity, we assume that the dilaton field is solely a function of time \cite{Ba}. If we narrow our analysis to an anisotropic but diagonal matrix form of the invariant metric $g_{ij}(t)$ \footnote{Note that one can also use another form of $g_{ij}$. For instance, in the case of four-dimensional Lie groups, one can refer to the results of \cite{Hervik1}.},
\renewcommand\theequation{\arabic{tempeq}\alph{equation}}
\setcounter{equation}{-1} \addtocounter{tempeq}{1}
\begin{eqnarray}
g_{ij}(t)=a_i^2(t) \delta_{ij},
\end{eqnarray}
where $a_i(t)$ represents scale factors. Then, the (0,0) and (i,i) components of Eq. (2a) take the following forms, respectively,
\cite{Gas,Mojaveri}
\renewcommand\theequation{\arabic{tempeq}\alph{equation}}
\setcounter{equation}{-1} \addtocounter{tempeq}{1}
\begin{eqnarray}
\dot{H_i}+H_i{\sum_{k=1}}^d H_k + H_i \dot{\phi}+V_i =0,
\end{eqnarray}
\renewcommand\theequation{\arabic{tempeq}\alph{equation}}
\setcounter{equation}{-1} \addtocounter{tempeq}{1}
\begin{eqnarray}
{\sum_{i=1}}^d(\dot{H_i}+ H_i^2)+\ddot{\phi}=0,
\end{eqnarray}
where the dot stands for the derivative with respect to $t$, and $H_i = \dot{a_i}/{a_i}$ represents the Hubble coefficients. The functions $V_k(a_i)$ are potential functions that depend on the type of Lie group. Additionally, the (i,0) components of Eq. (2a) imply constraints on the Hubble coefficients,\footnote{Here, it's important to note that there is no summation over $i$.}
\renewcommand\theequation{\arabic{tempeq}\alph{equation}}
\setcounter{equation}{-1} \addtocounter{tempeq}{1}
\begin{eqnarray}
{\sum_{k=1}}^d C_{ki}^k( H_i - H_k)=0,
\end{eqnarray}
where $C_{ij}^k$ represents the structure constants of the Lie algebras of the isometry groups. Note that, the components with $i\neq j$ in Eq. (2a) yield another set of constraint equations.
\renewcommand\theequation{\arabic{tempeq}\alph{equation}}
\setcounter{equation}{-1} \addtocounter{tempeq}{1}
\begin{eqnarray}
R_{i,j}=0,\qquad i\neq j.
\end{eqnarray}
Moreover, the dilaton equation (2c) takes the form
\renewcommand\theequation{\arabic{tempeq}\alph{equation}}
\setcounter{equation}{-1} \addtocounter{tempeq}{1}
\begin{eqnarray}
-2\ddot{\phi}-\dot{\phi}^2-2\dot{\phi}{\sum_{k=1}}^d H_k
-{\sum_{k=1}}^d V_k- ({\sum_{k=1}}^d H_k)^2
-{\sum_{k=1}}^d{H_k}^2-2{\sum_{k=1}}^d \dot{H_k}+\frac{H^2}{6}=0.
\end{eqnarray}
Thus, instead of solving Eqs. $(2a-2c)$, one can solve Eqs. $(6-10)$ to obtain homogeneous string cosmological models over space times with the isometry group $G$. For $4+1$-dimensional models, the isometry groups correspond to real four-dimensional Lie groups \cite{Hervik}. In our previous work \cite{La}, we focused on $4+1$-dimensional homogeneous anisotropic string models, considering cases with only a dilaton field and the absence of the $B$-field. Here, we will explore these models in the presence of the dilaton and antisymmetric matter ($B_{\mu\nu}\neq 0$). There are various classifications for four-dimensional Lie algebras \cite{Mu, Pet, Pat, Mac}. For our study, we adopt the classification provided by Patera and Winternitz \cite{Pat}. We express the mean radius $a$ in terms of scale factors $a_i$'s as presented in \cite{Ba, La}
\renewcommand\theequation{\arabic{tempeq}\alph{equation}}
\setcounter{equation}{-1} \addtocounter{tempeq}{1}
\begin{eqnarray}
a^4=a_1a_2a_3a_4.
\end{eqnarray}
Subsequently, by utilizing Eqs. (6), (7), along with (10), we obtain
\cite{Mojaveri}
\renewcommand\theequation{\arabic{tempeq}\alph{equation}}
\setcounter{equation}{-1} \addtocounter{tempeq}{1}
\begin{eqnarray}
\ddot{\phi}+\dot{\phi}^2+4\dot{\phi}\frac{\dot{a}}{a}+\frac{H^2}{6}=0,
\end{eqnarray}
Next, by introducing a new time coordinate $\tau$, \cite{Mojaveri}
\renewcommand\theequation{\arabic{tempeq}\alph{equation}}
\setcounter{equation}{-1} \addtocounter{tempeq}{1}
\begin{eqnarray}
d\tau=a^{-4} e^{-\phi}dt,
\end{eqnarray}
Equation (12) is expressed in the following form
\renewcommand\theequation{\arabic{tempeq}\alph{equation}}
\setcounter{equation}{-1} \addtocounter{tempeq}{1}
\begin{eqnarray}
\phi''+\frac{1}{6}H^2a^8e^{2\phi}=0,
\end{eqnarray}
where the prime stands for the derivative with respect to $\tau$.
Introducing this new time coordinate, Eqs. (7) may be rewritten as:
\renewcommand\theequation{\arabic{tempeq}\alph{equation}}
\setcounter{equation}{-1} \addtocounter{tempeq}{1}
\begin{eqnarray}
{\ln(a_i^2 e^{\phi})}''+2a^8 e^{2\phi}V_i-\phi'' =0.
\end{eqnarray}
On the other hand, Eq. (10) can be rewritten as the following initial value equation
\renewcommand\theequation{\arabic{tempeq}\alph{equation}}
\setcounter{equation}{-1} \addtocounter{tempeq}{1}
\begin{eqnarray}
\sum_{i<j}^4{\ln(a_i^2 e^{\phi})}'{\ln(a_j^2
e^{\phi})}'+2\sum_i^4V_i a^8
e^{2\phi}-2\sum_i^4\frac{{a_i}'}{a_i}\phi'-4{\phi'}^2-4\phi''-\frac{H^2}{3}a^8e^{2\phi}=0.
\end{eqnarray}
In this respect, it is sufficient to solve Eq. (15) with the constraints (8) and (9) and the initial values given in Eqs. (16). In Ref. \cite{Mojaveri}, we obtained metrics over four-dimensional Lie groups. For the self-containment of the paper, we have compiled the results in Appendix. It is important to note that to solve Eq. (15), we must determine the potentials $V_i(a_j)$. In the $4+1$-dimensional space-time, the potentials also depend on anti-symmetric matter fields. In the $4+1$-dimensional space-time, considering the condition $dH=0$, there are four possibilities for choosing the 3-form
$H=H_{\mu\nu\lambda}dx^{\mu}\wedge dx^{\nu}\wedge dx^{\lambda}$
\footnote {Here, we will consider only the cases in which $H_{ijk}$ represents the components of the 3-form on the space segment of $c(t)$ in space-time. Cases where $H$ is a 3-form over the entire space-time can be explored in a separate study.} \\

$\hspace{-4mm}\textbf{Case}\hspace{1mm} \textbf{i}:  \hspace{3mm}H=A
\hspace{1mm}
 (g^{-1}dg)^1\wedge(g^{-1}dg)^2\wedge(g^{-1}dg)^3,\:\:
\hspace{2mm}{}^{\star}db=e^{2\phi}dx^4\wedge dt,$\\
$\hspace{-4mm}\textbf{Case}\hspace{1mm}
\textbf{ii}:\hspace{2mm}H=A\hspace{1mm}
(g^{-1}dg)^1\wedge(g^{-1}dg)^2\wedge(g^{-1}dg)^4,\:\:
\hspace{3mm}{}^{\star}db=e^{2\phi}dx^{3}\wedge dt,$\\
$\hspace{-4mm}\textbf{Case}\hspace{1mm}
\textbf{iii}:\hspace{2mm}H=A\hspace{1mm}
(g^{-1}dg)^1\wedge(g^{-1}dg)^3\wedge(g^{-1}dg)^4,\:\:
\hspace{3mm}{}^{\star}db=e^{2\phi}dx^{2}\wedge dt,$\\
$\hspace{-4mm} \textbf{Case}\hspace{1mm}
\textbf{iv}:\hspace{2mm}H=A\hspace{1mm}
((g^{-1}dg)^2\wedge(g^{-1}dg)^3\wedge(g^{-1}dg)^4,\:\:
\hspace{3mm}{}^{\star}db=e^{2\phi}dx^{1}\wedge dt$.\\

\hspace{-8mm}
Here, $A$ is an arbitrary constant, and ${}^{\star}db$ represents the Hodge star of the three-form $db$, where the two-form $b$ is given by $b=B_{\mu\nu}dx^{\mu}\wedge dx^{\nu}$. In the next section, we will determine the potentials and attempt to solve eq (15) for all cases i-iv.
\vspace{-3mm}
\section{\bf Exact solutions}
\vspace{-1.5mm}
Here, for each case \textbf{i}-\textbf{iv} of $H$, we will construct the $(i,i)$ component of Eq. (2a) \cite{Mojaveri} over all real four-dimensional Lie groups of Appendix. In the following, by comparing these equations, we will obtain potentials $V_i$. Subsequently, we will attempt to solve Eqs. (15) to determine $a_i$ for $i=1, \ldots, 4$. Then, using (16), we can find the initial value equation.
\\
 We will perform this work for each of the real four-dimensional Lie algebras and for all the cases \textbf{i}-\textbf{iv} of $H$. The results are summarized in tables 1-4.

	\begin{sidewaystable}[H]
		\centering
		\begin{threeparttable}
			 \caption{$\hspace{-1mm}\textbf{Case}\hspace{1mm} \textbf{i}:  \hspace{3mm}H=A
\hspace{1mm}
 (g^{-1}dg)^1\wedge(g^{-1}dg)^2\wedge(g^{-1}dg)^3,\:\:
\hspace{2mm}{}^{\star}db=e^{2\phi}dx^4\wedge dt,$}
			\centering
			\begin{tabular}{| p{3em} | p{13em} | p{14em} | p{12em} | p{9.5em} |}
				\hline

                \multirow{2}{3em}{\scriptsize \tiny  Lie Algebra  }
				&\multirow{2}{5em} {\scriptsize  potentials }
				&\multirow{2}{5.3em} {\scriptsize equations }
				&\multirow{2}{9em} {\scriptsize  solution }
				&\multirow{2}{16em}{\scriptsize condition from eq (16)}
                \\
				
                &
			    &
				&
				&

				\\
				\hline
				
				\multirow{4}{3em}{\scriptsize $ I\oplus R$}
				
                &\multirow{4}{11em}{\scriptsize $V_1=V_2=V_3=-\frac{A^2}{2a_1^2a_2^2a_3^2},$}
                &\multirow{2}{8em}{\scriptsize$ \ln(a_c^2 e^{\phi})''=0,$}
                &\multirow{2}{8em}{\scriptsize $a_c^2e^{\phi}=p_ce^{N_c\tau},$}
				&\multirow{4}{8.5em}{\scriptsize $N_4^2+2(N_1+N_2+N_3)N_4+(N_2+N_3)N_1+N_2N_3-p_4^2=0.$}
				\\

				&\multirow{2}{8em}{\scriptsize$V_4=0.$}
				&\multirow{2}{8em}{\scriptsize$ ln(a_4^2 e^{\phi})''-\phi''=0,$}
				&\multirow{2}{8em}{\scriptsize $a_4^2e^{\phi}=\frac{p_4e^{N_4\tau}}{\cosh(p_4\tau)},$}
                &
				\\

				&
                &\multirow{2}{8em}{\scriptsize$ c= 1,2,3.$}
				&\multirow{1}{8em}{\scriptsize$ c= 1,2,3.$}
                &
				\\

				&
				&
				&
                &
				\\

				\hline
				
				\multirow{4}{3em}{\scriptsize $II\bigoplus R$ }

				&\multirow{2}{9em}{\scriptsize $V_1=\frac{a_1^2}{2a_2^2a_3^2}-\frac{A^2}{2a_1^2a_2^2a_3^2},$}
                &\multirow{2}{9em}{\scriptsize $\ln(a_1^2 e^{\phi})''+a_1^4a_4^2e^{2\phi}=0,$}
                &\multirow{2}{9em}{\scriptsize $a_1^2e^{\phi}=\frac{p_1\:e^{-N\tau}}{\cosh(p_4\tau)},$}
                &\multirow{3}{9em}{\scriptsize $-A^2p_1^2+p_4^2=0,$}
				\\

				&\multirow{2}{11em}{\scriptsize $V_2=V_3=-\frac{a_1^2}{2a_2^2a_3^2}-\frac{A^2}{2a_1^2a_2^2a_3^2}, $}
                &\multirow{1}{9em}{\scriptsize $\ln(a_2^2e^{\phi})''-a_1^4a_4^2e^{2\phi}=0,$}
                &\multirow{1}{9em}{\scriptsize $a_2^2e^{\phi}=p_2\cosh(p_4\tau)e^{\frac{p_2\tau}{2}},$}
                &\multirow{2}{8.5em}{\scriptsize $-2p_4^2+p_2(N+p_3)+Np_3=0.$}
				\\

				&
                &\multirow{1}{9em}{\scriptsize $\ln(a_3^2e^{\phi})''-a_1^4a_4^2e^{2\phi}=0,$}
                &\multirow{1}{9em}{\scriptsize $a_3^2e^{\phi}=p_3\cosh(p_4\tau)e^{\frac{p_3\tau}{2}},$}
				&
				
				\\

				&\multirow{1}{9em}{\scriptsize $V_4=0. $}
                &\multirow{1}{9em}{\scriptsize $\ln(a_4^2 e^{\phi})''-{\phi}''=0.$}
                &\multirow{1}{9em}{\scriptsize $a_4^2e^{\phi}=\frac{p_4\:e^{N\tau}}{\cosh(p_4\tau)}.$}
                &
				\\

                &
			    &
				&
				&

				\\

				\hline

				\multirow{5}{3em}{\scriptsize $ V\oplus R$}

				&\multirow{6}{12em}{\scriptsize $V_1=V_2=V_3=-\frac{2}{a_1^2}-\frac{A^2}{2a_1^2a_2^2a_3^2},$}
                &\multirow{2}{10em}{\scriptsize $\ln(a_1^2 e^{\phi})''-4a_2^2a_3^2a_4^2e^{2\phi}=0,$}
                &\multirow{2}{10em}{\scriptsize $
                a_1^2e^{\phi}=\frac{p_1\:e^{(N-p_3)\tau}}{2\sqrt{L_1}\cosh(p_4\tau)},$}
                &\multirow{3}{9.5em}{\scriptsize $-6N^2+2p_4^2-4p_2N-p_2^2=0,$}
				\\

				&\multirow{7}{9em}{\scriptsize$ V_4=0.$}
                &\multirow{1}{10em}{\scriptsize $\ln(a_2^2e^{\phi})''-4a_2^2a_3^2a_4^2e^{2\phi}=0,$}
                &\multirow{2}{9em}{\scriptsize $a_2^2e^{\phi}=\frac{L_2\:e^{(N+p_2)\tau}}{2\cosh(p_4\tau)},$}
                &\multirow{2}{8.5em}{\scriptsize $3p_1^2A^2+3p_4^2L_1=0$}.

                \\

                &
                &\multirow{1}{10em}{\scriptsize $\ln(a_3^2e^{\phi})''-4a_2^2a_3^2a_4^2e^{2\phi}=0,$}
				&\multirow{2}{10em}{\scriptsize $a_3^2e^{\phi}=\frac{p_1^2\:e^{(N-p_2-2p_3)\tau}}{2L_1L_2\cosh(p_4\tau)},$}
                &\multirow{2}{8.5em}{\scriptsize $p_3=2N.$}
				\\

				&
                &\multirow{2}{9em}{\scriptsize $\ln(a_4^2 e^{\phi})''-{\phi}''=0.$}
				&\multirow{3}{9em}{\scriptsize $a_4^2e^{\phi}=\frac{p_4e^{N\tau}}{cosh(p_4\tau)}.$}
                &
				\\

                &
                &\multirow{4}{13em}{\scriptsize  the constraint imposed by eqs (8) and (9) is $a_2a_3=B a_1^2.$}
				&
				&
				\\

                &
                &
				&
				&
				\\

                &
                &
				&
				&
				\\

				\hline

				\multirow{4}{3em}{\scriptsize	{$VI_{0}\oplus R$}}
				
               &\multirow{2}{12em}{\scriptsize $V_1=\frac{a_1^2}{2a_2^2a_3^2}-\frac{a_2^2}{2a_1^2a_3^2}-\frac{A^2}{2a_1^2a_2^2a_3^2},$}
                &\multirow{2}{14em}{\scriptsize $\ln(a_1^2 e^{\phi})''+(a_1^4-a_2^4)a_4^2e^{2\phi}=0,$}
                &\multirow{3}{10em}{\scriptsize $a_1^2 e^{\phi}=a_2^2e^{\phi}=\sqrt{L_1}\:e^{\frac{p_1}{2}\tau},$}
                &\multirow{4}{8.5em}{\scriptsize $ \frac{p_1^2}{4}-p_4^2=0.$}
				\\

                &\multirow{2}{12em}{\scriptsize $V_2=-\frac{a_1^2}{2a_2^2a_3^2}+\frac{a_2^2}{2a_1^2a_3^2}-\frac{A^2}{2a_1^2a_2^2a_3^2},$}
                &\multirow{1}{14em}{\scriptsize $\ln(a_2^2e^{\phi})''-(a_1^4-a_2^4)a_4^2e^{2\phi}=0,$}
				&
				&\multirow{4}{8.5em}{\scriptsize $ L_1=L_2,$}
				\\

				&\multirow{3}{13em}{\scriptsize $V_3=-\frac{a_1^2}{2a_2^2a_3^2}-\frac{a_2^2}{2a_1^2a_3^2}-\frac{1}{a_3^2}-\frac{A^2}{2a_1^2a_2^2a_3^2},$}
                &\multirow{3}{14em}{\scriptsize $\ln(a_3^2e^{\phi})''-(a_1^4+a_2^4)a_4^2e^{2\phi}-2a_1^2a_2^2a_4^2e^{2\phi}=0,$}
                &\multirow{2}{14em}{\scriptsize $a_3^2e^{\phi}=exp\left(\frac{4A^2L_1L_2}{p_1^2}\,e^{(p_1+2N)\tau}\right).$}
                &
				
				\\

				&\multirow{4}{10em}{\scriptsize $V_4=0.$}
                &\multirow{5}{14em}{\scriptsize $\ln(a_4^2 e^{\phi})''-\phi''=0,$}
				&\multirow{4}{12em}{\scriptsize $a_4^2e^{\phi}=\frac{p_4e^{N\tau}}{\cosh(p_4\tau)}.$}
				&
				\\

                &
                &
                &
                &
                \\

                &
                &
                &
                &
                \\

				\hline
			\end{tabular}

\caption*{\scriptsize Here for all Lie algebra of this table $H_{123}=A $  and  $\phi=\ln(\frac{p_4}{A^2})-N\tau-\ln(\cosh(p_4\tau)).$}

			\label{table:1}
		\end{threeparttable}
		
	\end{sidewaystable}

	\newpage
	
	\begin{sidewaystable}[H]
		\centering
		\begin{threeparttable}

			 \caption*{ $\text{Table 1}:\hspace{-1mm}\textbf{Case}\hspace{1mm} \textbf{i}:continued $}
			
			\begin{tabular}{| p{3em} | p{12.5em} | p{16em} | p{10.8em} | p{8.5em} | }
				\hline
				
				\multirow{2}{3em}{\scriptsize\tiny   Lie Algebra  }
				&\multirow{2}{5em} {\scriptsize  potentials }
				&\multirow{2}{5.3em} {\scriptsize equations }
				&\multirow{2}{9em} {\scriptsize  solution }
				&\multirow{2}{16em}{\scriptsize condition from eq (16)}
				\\
				
				&
			    &
				&
				&
				\\
				\hline
				
				\multirow{4}{3em}{\scriptsize $VII_0\bigoplus R$}
				&\multirow{2}{11em}{\scriptsize $V_1=\frac{a_1^2}{2a_2^2a_3^2}-\frac{a_2^2}{2a_1^2a_3^2}-\frac{A^2}{2a_1^2a_2^2a_3^2},$}
				&\multirow{2}{11em}{\scriptsize $\ln(a_1^2 e^{\phi})''+(a_1^4-a_2^4)a_4^2e^{2\phi}=0,$}
				&\multirow{2}{11em}{\scriptsize $a_1^2e^{\phi}=a_2^2e^{\phi}=\sqrt{L_1}\:e^{\frac{p_1}{2}\tau},$}
				&\multirow{5}{9em}{\scriptsize $p_1^2+(8N+4p_3)p_1+4N^2+8p_3N-4p_4^2=0.$}

				\\

		        &\scriptsize $V_2=-\frac{a_1^2}{2a_2^2a_3^2}+\frac{a_2^2}{2a_1^2a_3^2}-\frac{A^2}{2a_1^2a_2^2a_3^2},$
				&\scriptsize $\ln(a_2^2e^{\phi})''-(a_1^4-a_2^4)a_4^2e^{2\phi}=0,$
				&\scriptsize $a_3^2e^{\phi}=L_2\:e^{p_3\tau},$
				&
				\\

				&\scriptsize $V_3=-\frac{a_1^2}{2a_2^2a_3^2}-\frac{a_2^2}{2a_1^2a_3^2}+\frac{1}{a_3^2}-\frac{A^2}{2a_1^2a_2^2a_3^2},$
				&\scriptsize $\ln(a_3^2e^{\phi})''-(a_1^4+a_2^4)a_4^2e^{2\phi}+2a_1^2a_2^2a_4^2e^{2\phi}=0,$
				&\scriptsize $a_4^2e^{\phi}=\frac{p_4\:e^{N\tau}}
{\cosh(p_4\tau)}.$
				&
				\\
				
				&\scriptsize $V_4=0.$
				&\scriptsize $\ln(a_4^2 e^{\phi})''-\phi''=0.$
				&
				&
				\\

                &
                &
                &
                &
               \\

				\hline
				
				\multirow{4}{3em}{\scriptsize
$VII_b\oplus R$ }

				&\multirow{4}{11em}{\scriptsize $V_1=-\frac{a_3^2}{2a_1^2a_2^2}-\frac{a_2^2}{2a_1^2a_3^2}-\frac{2b^2-1}{a_1^2}-\frac{A^2}{2a_1^2a_2^2a_3^2},$}
				&\multirow{3}{16em}{\scriptsize $\ln(a_1^2 e^{\phi})''-\left((a_2^2-a_3^2)^2+4b^2a_2^2a_3^2\right)a_4^2e^{2\phi}=0,$}
				&\multirow{4}{11em}{\scriptsize $a_1^2e^{\phi}=\frac{p_1\:e^{(N-p_1)\tau}}{2bB\sqrt{L_1}\sinh(p_4\tau)},$}
                &\multirow{5}{8em}{\scriptsize $-A^2p_1^2+L_1^2=0.$}
				\\

				&\multirow{4}{13em}{\scriptsize $V_2=\frac{a_2^2}{2a_1^2a_3^2}-\frac{a_3^2}{2a_1^2a_2^2}-\frac{2b^2}{a_1^2}-\frac{A^2}{2a_1^2a_2^2a_3^2},$}
				&\multirow{2}{15em}{\scriptsize $\ln(a_2^2e^{\phi})''-4b^2a_2^2a_3^2a_4^2e^{2\phi}=0,$}
				&\multirow{4}{13em}{\scriptsize $a_2^2e^{\phi}=a_3^2e^{\phi}=\frac{p_1\:e^{(N-p_1)\tau}}{2b\sqrt{L_1}\sinh(p_4\tau)},$}
                &
				\\

				&\multirow{5}{11em}{\scriptsize $V_3=-\frac{a_2^2}{2a_1^2a_3^2}+\frac{a_3^2}{2a_1^2a_2^2}-\frac{2b^2}{2a_1^2},$}
				&\multirow{3}{15em}{\scriptsize $\ln(a_3^2e^{\phi})''-4b^2a_2^2a_3^2a_4^2e^{2\phi}=0,$}
				&\multirow{5}{13em}{\scriptsize $a_4^2e^{\phi}=\frac{p_4e^{N\tau}}{\cosh(p_4\tau)}.$}
                &
				\\

				&\multirow{6}{11em}{\scriptsize $ V_4=0.$}
				&\multirow{4}{11em}{\scriptsize $\ln(a_4^2 e^{\phi})''-\phi''=0.$}

				&
                &
				\\

                &
				&\multirow{4}{16em}{\scriptsize conditions from eqs (8) and (9) $a_2^2=a_3^2$, $a_2a_3=D a_1^2,$ }
				&
				&
				\\

                &
				&
				&
				&
				\\
				
				&
				&
				&
				&
				\\

                &
                &
                &
                &
               \\

				\hline

				\multirow{5}{3em}{\scriptsize $ A_{4,1}$}
				
                &\multirow{2}{10em}{\scriptsize $V_1=\frac{a_1^2}{2a_2^2a_4^2}-\frac{A^2}{a_1^2a_2^2a_3^2},$}
				&\multirow{3}{14em}{\scriptsize $\ln(a_1^2 e^{\phi})''+a_1^4a_3^2e^{2\phi}=0,$}
				&\multirow{2}{10em}{\scriptsize $\phi=-\frac{D^2\tau^2}{2}+F\tau,$}
				&\multirow{5}{9em}{\scriptsize{\tiny $3FC_1-2C_1^2-C_1C_2-F^2-FC_2-3D^2=0.$}}
				\\

				&\multirow{2}{13em}{\scriptsize $V_2=-\frac{a_1^2}{2a_2^2a_4^2}+\frac{a_2^2}{2a_3^2a_4^2}-\frac{A^2}{a_1^2a_2^2a_3^2},$}
                &\multirow{3}{16em}{\scriptsize $\ln(a_2^2e^{\phi})''+\left(a_1^2a_2^4-a_3^2a_1^4\right)e^{2\phi}=0,$}
                &\multirow{1}{13em}{\scriptsize $a_1^2e^{\phi}=p_1e^{b_1\tau^2+C_1\tau},$}
				&
				\\

				&\multirow{3}{10em}{\scriptsize $V_3=-\frac{a_2^2}{2a_3^2a_4^2}-\frac{A^2}{a_1^2a_2^2a_3^2},$}
                &\multirow{3}{16em}{\scriptsize $ \ln(a_3^2e^{\phi})''-a_1^2a_2^4e^{2\phi}=0,$}
			    &\multirow{1}{13em}{\scriptsize $a_2^2e^{\phi}=p_2e^{C_2\tau},$}
				&
				\\

				&\multirow{4}{10em}{\scriptsize $V_4=-\frac{a_1^2}{2a_2^2a_4^2}-\frac{a_2^2}{2a_3^2a_4^2}.$}
                &\multirow{4}{16em}{\scriptsize $\ln(a_4^2e^{\phi})''-\left(a_1^2a_2^4+a_3^2a_1^4\right)e^{2\phi}-\phi''=0.$}
                &\multirow{1}{13em}{\scriptsize $a_3^2e^{\phi}=p_3e^{\frac{D^2\tau^2}{2}+(F-2C_1)\tau+G},$}	
				&
				\\
				
				&
				&
				&\multirow{1}{13em}{\scriptsize $a_4^2e^{\phi}=B^2e^{\frac{D^2\tau^2}{2}-(F-2C_1)\tau-G}.$}
				&
				\\

                &
                &
                &
                &
               \\

				\hline

				\multirow{4}{3em}{\scriptsize	$A_{4,5}^{(a,b)}$}
				
                &\multirow{6}{10em}{\scriptsize $V_1=V_2=V_3= -\frac{A^2}{2a_1^2a_2^2a_3^2}$,}
				&\multirow{3}{10em}{\scriptsize $\ln(a_1^2 e^{\phi})''=0,$}
				&\multirow{3}{10em}{\scriptsize $a_1^2e^{\phi}=p_1^2e^{N_1\tau},$}
				&\multirow{3}{8em}{\scriptsize $N_1N_2+N_1N_3+N_2N_3+8p_4^2-N_4^2=0,$}
				\\

				&
				&\multirow{3}{10em}{\scriptsize $\ln(a_2^2e^{\phi})''=0,$}
				&\multirow{3}{10em}{\scriptsize $a_2^2e^{\phi}=p_2^2e^{N_2\tau},$}
				&\multirow{2}{8em}{\scriptsize $N_1+N_2+N_3=N_4.$}
				\\

				&
				&\multirow{4}{10em}{\scriptsize $\ln(a_3^2e^{\phi})''=0,$}
				&\multirow{4}{10em}{\scriptsize $a_3^2e^{\phi}=p_3^2e^{N_3\tau},$}
				&
				\\

				&\multirow{3}{10em}{\scriptsize $V_4=-\frac{a^2+b^2+1}{a_4^2}.$}
				&\multirow{5}{16em}{\scriptsize $\ln(a_4^2e^{\phi})''-2\left(a^2+b^2+1\right)a_1^2a_2^2a_3^2\:e^{2\phi}-\phi''=0.$}
				&\multirow{5}{10em}{\scriptsize $a_4^2e^{\phi}=\frac{2p_4e^{-N_4\tau}}{A\sinh^4(p_4\tau)}.$}
				&
				\\

                &
                &
                &
                &
                \\

                &
                &
                &
                &
                \\

                &
                &
                &
                &
                \\

				\hline
			\end{tabular}
\caption*{\scriptsize  Here for all Lie algebras of this table except  $A_{4,1}$,\hspace{1mm}  $\phi=\ln(\frac{p_4}{A^2})-N\tau-\ln(\cosh(p_4\tau))$;  and $H_{123}=A $ .}
	
		\end{threeparttable}
		
	\end{sidewaystable}
	
	\newpage
	
	\begin{sidewaystable}[H]
		
		\begin{threeparttable}
			 \caption*{$\text{Table 1}:\hspace{-1mm}\textbf{Case}\hspace{1mm} \textbf{i}:continued$}
			\centering
			\begin{tabular}{| p{3em} | p{11.8em} | p{5.5em} | p{17.5em} | p{6.5em} | p{5em} | }
				\hline
				
				\multirow{2}{5.3em}{\scriptsize\tiny Lie Algebra  }
				&\multirow{2}{5em} {\scriptsize  potentials }
                &\multirow{2}{5em}{\scriptsize $H_{\mu\nu\lambda}$}
				&\multirow{2}{5.3em} {\scriptsize equations }
				&\multirow{2}{6.5em} {\scriptsize  solution }
				&\multirow{2}{5em}{\scriptsize\tiny condition of eq (16)}
				\\
				
				&
                &
			    &
				&
				&
				\\
				\hline
				
				\multirow{4}{5.3em}{\scriptsize $A_{4,5}^{(a,a)}$}
				&\multirow{6}{11em}{\scriptsize $V_1=V_2=V_3=-\frac{A^2}{2a_1^2a_2^2a_3^2},$}
                &\multirow{4}{11em}{\scriptsize $ H_{123}=A\:e^{(1+2a)\:x_4},$}
				&\multirow{2}{14em}{\scriptsize $\ln(a_1^2 e^{\phi})''=0,$}
				&\multirow{2}{11em}{\scriptsize $a_1^2e^{\phi}=p_1^2e^{N_1\tau},$}
				&\multirow{3}{5em}{\scriptsize\tiny $N_1N_2+N_1N_3+N_2N_3+8p_4^2-N_4^2=0,$}

				\\

		        &
                &\multirow{4}{8em}{\scriptsize \tiny $dH=0$ with condition $2a=-1,$}
				&\multirow{2}{14em}{\scriptsize $\ln(a_2^2e^{\phi})''=0,$}
				&\multirow{1}{8em}{\scriptsize $a_2^2e^{\phi}=p_2^2e^{N_2\tau},$}
				&\multirow{4}{5em}{\scriptsize\tiny $N_1+N_2+N_3=N_4.$}
				\\

				&
                &
				&\multirow{3}{14em}{\scriptsize $\ln(a_3^2e^{\phi})''=0,$}
				&\multirow{2}{8em}{\scriptsize $a_3^2e^{\phi}=p_3^2e^{N_3\tau},$}
				&
				\\
				
				&\multirow{2}{8em}{\scriptsize $V_4=-\frac{2a^2+1}{a_4^2}.$}
                &
				&\multirow{4}{17.5em}{\scriptsize $\ln(a_4^2e^{\phi})''-\left(2a^2+1\right)a_1^2a_2^2a_3^2\:e^{2\phi}-\phi''=0.$}
				&
				&
				\\

                &
                &
				&
				&\multirow{1}{10em}{\scriptsize $a_4^2e^{\phi}=\frac{2p_4e^{N_4\tau}}{A\sinh^4(p_4\tau)}.$}
                &
				\\

                &
                &
				&
				&
                &
				\\

				\hline
				
				\multirow{4}{5.3em}{\scriptsize $A_{4,5}^{(a,1)}$}

				&\multirow{4}{11em}{\scriptsize $V_1=V_2=V_3=-\frac{A^2}{2a_1^2a_2^2a_3^2},$}
                &\multirow{4}{11em}{\scriptsize $ H_{123}=A\:e^{(2+a)\:x_4},$}
				&\multirow{2}{11em}{\scriptsize $\ln(a_1^2 e^{\phi})''=0,$}
				&\multirow{2}{11em}{\scriptsize $a_1^2e^{\phi}=p_1^2e^{N_1\tau},$}
                &\multirow{3}{5em}{\scriptsize\tiny $ N_1N_2+N_1N_3+N_2N_3+8p_4^2-N_4^2=0,$}
				\\

				&
                &\multirow{4}{8em}{\scriptsize\tiny $dH=0$ with condition $a=-2.$}
				&\multirow{2}{10em}{\scriptsize $\ln(a_2^2e^{\phi})''=0,$}
				&\multirow{2}{8em}{\scriptsize $a_2^2e^{\phi}=p_2^2e^{N_2\tau},$}
                &\multirow{4}{5em}{\scriptsize\tiny $N_1+N_2+N_3=N_4.$}
				\\

				&
                &\multirow{2}{8em}{\scriptsize }
				&\multirow{3}{8em}{\scriptsize $\ln(a_3^2e^{\phi})''=0,$}
				&\multirow{3}{8em}{\scriptsize $a_3^2e^{\phi}=p_3^2e^{N_3\tau},$}
                &
				\\

				&\scriptsize $V_4=-\frac{a^2+2}{a_4^2}.$
                &
				&\multirow{4}{17.5em}{\scriptsize $\ln(a_4^2e^{\phi})''-\left(a^2+2\right)a_1^2a_2^2a_3^2\:e^{2\phi}-\phi''=0.$}
				&\multirow{4}{10em}{\scriptsize $a_4^2e^{\phi}=\frac{2p_4e^{-N_4\tau}}{A\sinh^4(p_4\tau)}.$}
                &
				\\

                &
                &
				&
				&
                &
				\\

                &
                &
				&
				&
                &
				\\

				\hline

				\multirow{6}{5.3em}{\scriptsize $A_{4,6}^{(a,b)}$}
				
                &\multirow{2}{11em}{\scriptsize $V_1=-\frac{A^2}{2a_1^2a_2^2a_3^2},$}
                &\multirow{4}{11em}{\scriptsize $ H_{123}=A\:e^{(a+2b)x_4},$ }
				&\multirow{2}{17.5em}{\scriptsize $\ln(a_1^2 e^{\phi})''-2\left(a^2+2ab\right)a_1^2a_2^2a_3^2e^{2\phi}=0,$}
				&\multirow{2}{6.5em}{\scriptsize $a_1^2e^{\phi}=p_1^2e^{N_1\tau},$}
				&\multirow{2}{6em}{\scriptsize $6p_4^3=12Ab^2p_1^6$.}
				\\

				&\multirow{2}{14em}{\scriptsize $V_2=\frac{a_2^2}{2a_3^2a_4^2}-\frac{a_3^2}{2a_2^2a_4^2}-\frac{A^2}{2a_1^2a_2^2a_3^2},$}
                &\multirow{4}{8em}{\scriptsize\tiny $dH=0,$ with condition $a=-2b,$}
				&\multirow{2}{17.5em}{\scriptsize $\ln(a_2^2e^{\phi})''-a_1^2\left(a_3^4-a_2^4+2b(2b^2+a)a_2^2a_3^2\right)e^{2\phi}=0,$}
				&\multirow{1}{6em}{\scriptsize $a_2^2e^{\phi}=p_2^2e^{N_2\tau},$}
				&\multirow{2}{6em}{\scriptsize $N^2=4p^4.$}
				\\
				
				&\multirow{3}{14em}{\scriptsize $V_3=-\frac{a_2^2}{2a_3^2a_4^2}+\frac{a_3^2}{2a_2^2a_4^2}-\frac{A^2}{2a_1^2a_2^2a_3^2},$}
                &
                &\multirow{4}{17.5em}{\scriptsize $\ln(a_3^2e^{\phi})''-a_1^2\left(a_2^4-a_3^4+2b(2b^2+a)a_2^2a_3^2\right)e^{2\phi}=0,$}
                &\multirow{1}{6em}{\scriptsize $a_3^2e^{\phi}=p_3^2e^{N_3\tau},$}
                &
				
				\\

				&\multirow{4}{12em}{\scriptsize $V_4=-\frac{a_2^2}{2a_3^2a_4^2}-\frac{a_3^2}{2a_2^2a_4^2}-\frac{(a^2+2b^2-1)}{a_4^2}.$}
                &
                &
                \multirow{6}{17.5em}{\scriptsize $\ln(a_4^2e^{\phi})''-a_1^2\left(a_2^4+a_3^4+2(2b^2+a^2-1)a_2^2a_3^2\right)e^{2\phi}-\phi''=0.$}
                &\multirow{1}{6em}{\scriptsize $a_4^2e^{\phi}=\frac{2p_4e^{-N_4\tau}}{A\sinh^4(p_4\tau)}.$}
                &
				\\

                &
                &
				&
				&
                &
				\\
                &
                &
				&
				&
                &
				\\
                &
                &
				&
				&
                &
				\\

				\hline
			\end{tabular}
	
			\caption*{\scriptsize Here for all Lie algebras of this table  $\phi=\ln(\frac{p_4}{A^2})-N\tau-\ln(\cosh(p_4\tau)).$}
		\end{threeparttable}
		
	\end{sidewaystable}

	\newpage

	\begin{sidewaystable}[H]
		
		\begin{threeparttable}
			 \caption*{$\text{Table 1}: \hspace{1mm}\textbf{Case}\hspace{1mm} \textbf{i}:$ continued}
			\centering
			\begin{tabular}{| p{2.3em} | p{13.5em} | p{7.5em} | p{17em} | p{5em} | }
\hline
\multirow{2}{2.3em}{\scriptsize \tiny  Lie Algebra  }
				&\multirow{2}{5em} {\scriptsize  potentials }
                &\multirow{2}{5em}{\scriptsize $H_{\mu\nu\lambda}$}
				&\multirow{2}{5.3em} {\scriptsize equations }
				&\multirow{2}{6.5em} {\scriptsize  solution }

				\\
				
				&
                &
			    &
				&

				\\
				\hline
				
				\multirow{2}{5.3em}{\scriptsize $A_{4,8}$ }
				&\multirow{2}{11em}{\scriptsize $V_1=\frac{a_1^2}{2a_2^2a_3^2}-\frac{A^2}{2a_1^2a_2^2a_3^2},$}
                &\multirow{4}{11em}{\scriptsize $H_{123}=A\:e^{2b\:x_4},$}
				&\multirow{2}{11em}{\scriptsize $\ln(a_1^2 e^{\phi})''+a_1^4a_4^2e^{2\phi}=0,$}
				&\multirow{5}{5em}{\scriptsize\tiny We could not find a solution}
				
				\\
				
				&\multirow{3}{11em}{\scriptsize $V_2=-\frac{a_1^2}{2a_2^2a_3^2}-\frac{A^2}{2a_1^2a_2^2a_3^2},$}
                &
			    &\multirow{2}{11em}{\scriptsize $\ln(a_2^2e^{\phi})''-a_1^4a_4^2e^{2\phi}=0,$}
				&
				
				\\

                &\multirow{4}{11em}{\scriptsize $V_3=V_2,$}
                &
			    &\multirow{3}{11em}{\scriptsize $\ln(a_3^2e^{\phi})''-a_1^4a_4^2e^{2\phi}=0,$}
				&
				
				\\

                &\multirow{5}{10em}{\scriptsize $V_4=-\frac{2}{a_4^2}.$}
                &
			    &\multirow{4}{12em}{\scriptsize $\ln(a_4^2e^{\phi})''-4a_1^2a_2^2a_3^2e^{2\phi}-\phi''=0.$}
				&
				\\

               &
               &
               &\multirow{5}{16em}{\scriptsize eq (8) impose the  restriction:}
               &
               \\

               &
               &
               &\multirow{5}{14em}{\scriptsize $a_2=Ba_3,$}
               &
               \\

                &
                &
			    &
				&
				
				\\
                &
                &
			    &
				&

				\\

				\hline
				
				\multirow{4}{5.3em}{\scriptsize $A_{4,9}^{b}$}
				&\multirow{2}{10em}{\scriptsize $V_1=\frac{a_1^2}{2a_2^2a_3^2}-\frac{8}{a_4^2}-\frac{A^2}{2a_1^2a_2^2a_3^2},$}
                &\multirow{4}{7.5em}{\scriptsize $H_{123}=A\:e^{(3+b)\:x_4},$}
				&\multirow{2}{14em}{\scriptsize $\ln(a_1^2 e^{\phi})''+\left(a_1^4a_4^2-16a_1^2a_2^2a_3^2\right)e^{2\phi}=0,$}
                &\multirow{4}{5em}{\scriptsize\tiny We could not find a solution}
				
				\\

				&\multirow{2}{11em}{\scriptsize $V_2=-\frac{a_1^2}{2a_2^2a_3^2}+\frac{4}{a_4^2}-\frac{A^2}{2a_1^2a_2^2a_3^2},$}
                &\multirow{5}{8em}{\scriptsize$dH=0$,with condition $b=-3,$}
				&\multirow{2}{14em}{\scriptsize $\ln(a_2^2e^{\phi})''-\left(a_1^4a_4^2-8a_1^2a_2^2a_3^2\right)e^{2\phi}=0,$}
				&
				
				\\
				
				&\multirow{3}{11em}{\scriptsize $V_3=-\frac{a_1^2}{2a_2^2a_3^2}-\frac{12}{a_4^2}-\frac{A^2}{2a_1^2a_2^2a_3^2},$}
                &
				&\multirow{3}{14em}{\scriptsize $\ln(a_3^2e^{\phi})''-(a_1^4a_4^2+24a_1^2a_2^2a_3^2)e^{2\phi}=0,$}
				&
				
				\\
				
				&\multirow{4}{11em}{\scriptsize$ V_4=-\frac{14}{a_4^2}.$}
                &
				&\multirow{4}{14em}{\scriptsize $ \ln(a_4^2e^{\phi})''-28a_1^2a_2^2a_3^2e^{2\phi}-\phi''=0.$}
				&
				
				\\

                &
                &
			    &\multirow{5}{16em}{\scriptsize eq (8) impose restriction:}
				&

				\\

                &
                &
			    &\multirow{5}{8em}{\scriptsize $a_1^{-2}a_2a_3^{-3}=Ba_4^{-4},$}
				&

				\\

                &
                &
			    &
				&

				\\

                &
                &
			    &
				&

				\\

				\hline
				
				\multirow{4}{5.3em}{\scriptsize $A_{4,10}$}

				&\multirow{2}{9em}{\scriptsize $V_1=-\frac{a_1^2}{2a_2^2a_3^2}-\frac{A^2}{2a_1^2a_2^2a_3^2},$}
                &\multirow{2}{9em}{\scriptsize $H_{123}=A$}
				&\multirow{2}{9em}{\scriptsize $\ln(a_1^2 e^{\phi})''-a_1^4a_4^2e^{2\phi}=0,$}
                &\multirow{4}{5em}{\scriptsize\tiny We could not find a solution}
				
				\\

				&\multirow{2}{14em}{\scriptsize $V_2=\frac{a_1^2}{2a_2^2a_3^2}-\frac{a_2^2}{2a_3^2a_4^2}+\frac{a_3^2}{2a_2^2a_4^2}-\frac{A^2}{2a_1^2a_2^2a_3^2},$}
                &
				&\multirow{2}{14.8em}{\scriptsize $\ln(a_2^2e^{\phi})''+\left(a_1^2a_3^4-a_1^2a_2^4+a_4^2a_1^4\right)e^{2\phi}=0,$}
				&
				\\

				&\multirow{3}{14em}{\scriptsize $V_3=\frac{a_1^2}{2a_2^2a_3^2}+\frac{a_2^2}{2a_3^2a_4^2}-\frac{a_3^2}{2a_2^2a_4^2}-\frac{A^2}{2a_1^2a_2^2a_3^2},$}
                &
				&\multirow{3}{14.8em}{\scriptsize $\ln(a_3^2e^{\phi})''+\left(a_1^2a_2^4-a_1^2a_3^4+a_4^2a_1^4\right)e^{2\phi}=0,$}
				&
				\\

				&\multirow{4}{14em}{\scriptsize $V_4=\frac{a_2^2}{2a_3^2a_4^2}+\frac{a_3^2}{2a_2^2a_4^2}-\frac{1}{a_4^2}.$}
                &
				&\multirow{4}{17.5em}{\scriptsize $\ln(a_4^2e^{\phi})''+\left(a_1^2a_3^4-2a_1^2a_2^2a_3^2+a_1^2a_2^4\right)e^{2\phi}-\phi''=0.$}
				&
				\\

                &
                &
			    &
				&

				\\

                &
                &
			    &
				&

				\\

				\hline

				\multirow{6}{5.3em}{\scriptsize $A_{4,11}^{a}$}
				
                &\multirow{2}{10em}{\scriptsize $V_1=\frac{a_1^2}{2a_2^2a_3^2}-\frac{A^2}{2a_1^2a_2^2a_3^2},$}
                &\multirow{2}{10em}{\scriptsize $H_{123}=A.$}
				&\multirow{2}{16em}{\scriptsize$ \ln(a_1^2 e^{\phi})''+a_1^4a_4^2\:e^{2\phi}=0,$}
				&\multirow{4}{5em}{\scriptsize\tiny  We could not find a solution}
				\\

				&\multirow{2}{10em}{\scriptsize $V_2=-\frac{a_1^2}{2a_2^2a_3^2}-\frac{A^2}{2a_1^2a_2^2a_3^2},$}
                &\multirow{2}{9em}{\scriptsize $dH=0$ with condition $a=0.$}
				&\multirow{2}{16em}{\scriptsize$ \ln(a_2^2e^{\phi})''-a_4^2a_1^4\:e^{2\phi}=0,$}
				&
				\\

                &\multirow{3}{10em}{\scriptsize $V_3=V_2,$}
                &
				&\multirow{2}{16em}{\scriptsize$ \ln(a_3^2e^{\phi})''-a_4^2a_1^4\:e^{2\phi}=0,$}
				&
				
				\\

				&\multirow{4}{12em}{\scriptsize $V_4=-\frac{a_2^2}{2a_3^2a_4^2}-\frac{a_3^2}{2a_2^2a_4^2}+\frac{1}{a_4^2}$.}
                &
				&\multirow{3}{16em}{\scriptsize $\ln(a_4^2e^{\phi})''-\phi''=0.$}
				&
				\\

                &
                &
				&\multirow{4}{14em}{\scriptsize The eq (8) led to the constraint:}
                &
                \\

                &
                &
				&\multirow{4}{9em}{\scriptsize $a_2^2=a_3^2.$}
                &
			    \\

                &
                &
				&
                &
			    \\

                &
                &
				&
                &
			    \\

				\hline
			\end{tabular}
\caption*{\scriptsize Here for all Lie algebra of this table $\phi=\ln(\frac{p_4}{A^2})-N\tau-\ln(\cosh(p_4\tau)).$}

		\end{threeparttable}
		
	\end{sidewaystable}

	\begin{sidewaystable}[H]
		
		\begin{threeparttable}
			 \caption{$\hspace{-1mm}\textbf{Case}\hspace{1mm} \textbf{ii}:  \hspace{3mm}H=A
\hspace{1mm}
 (g^{-1}dg)^1\wedge(g^{-1}dg)^2\wedge(g^{-1}dg)^4,\:\:
\hspace{2mm}{}^{\star}db=e^{2\phi}dx^3\wedge dt,$}
			\centering
			\begin{tabular}{| p{3em} | p{11em} | p{18em} | p{9em} | p{9em} | }
				\hline
				
				\multirow{2}{3em}{\scriptsize\tiny  Lie Algebra  }
				&\multirow{2}{5em} {\scriptsize  potentials }
				&\multirow{2}{5.3em} {\scriptsize equations }
				&\multirow{2}{9em} {\scriptsize  solution }
				&\multirow{2}{16em}{\scriptsize condition from eq (16)}
				\\
				
				&
			    &
				&
				&
				\\
				\hline
				
				\multirow{4}{3em}{\scriptsize $ I\oplus R$}
				&\multirow{3}{11em}{\scriptsize $V_1=V_2=V_4=-\frac{A^2}{2a_1^2a_2^2a_4^2},$}
				&\multirow{2}{8em}{\scriptsize$\ln(a_3^2 e^{\phi})''-\phi''=0,$}
				&\multirow{2}{8em}{\scriptsize $a_3^2e^{\phi}=\frac{p_4e^{N_3\tau}}{\cosh(p_4\tau)},$}
				&\multirow{4}{8em}{\scriptsize $N_3^2+2(N_1+N_2+N_4)N_3+(N_2+N_4)N_1+N_2N_4-p_4^2=0.$}

				\\

				&\multirow{2}{8em}{\scriptsize$ V_3=0.$}
				&\multirow{1}{8em}{\scriptsize $\ln(a_c^2 e^{\phi})''=0, $}
				&\multirow{2}{8em}{\scriptsize $a_c^2e^{\phi}=p_ce^{N_c\tau}.$}
                &
				\\

				&
				&\multirow{2}{8em}{\scriptsize$c= 1,2,4,$}
				&\multirow{2}{8em}{\scriptsize$c= 1,2,4,$}
				&
				\\
				
				&
				&
				&
				&
				\\

				\hline
				
				\multirow{4}{3em}{\scriptsize $II\bigoplus R$ }

				&\multirow{2}{9em}{\scriptsize $V_1=\frac{a_1^2}{2a_2^2a_3^2}-\frac{A^2}{2a_1^2a_2^2a_4^2},$}
				&\multirow{2}{9em}{\scriptsize $\ln(a_1^2 e^{\phi})''+a_1^4a_4^2e^{2\phi}=0,$}
				&\multirow{3}{11
em}{\scriptsize $a_1^2e^{\phi}=\frac{p_1\:e^{\frac{N\tau}{4}}}{\cosh^2(p_5\tau)},$}
                &\multirow{3}{9em}{\scriptsize $ -6N^2-5Np_2-64p_5^2=0,$}
				\\

				&\multirow{1}{9em}{\scriptsize $V_2=-\frac{a_1^2}{2a_2^2a_3^2}-\frac{A^2}{2a_1^2a_2^2a_4^2},$}
				&\multirow{1}{11em}{\scriptsize $\ln(a_2^2e^{\phi})''-a_1^4a_4^2e^{2\phi}=0,$}
				&\multirow{2}{11em}{\scriptsize $a_2^2e^{\phi}=p_1\cosh^2(p_5\tau)e^{\frac{p_2\tau}{4}},$}
                &\multirow{2}{9em}{\scriptsize $ -p_4p_1^2+2p_5^2=0.$}
				\\

				&\multirow{2}{9em}{\scriptsize $V_3=-\frac{a_1^2}{2a_2^2a_3^2},$}
				&\multirow{2}{11em}{\scriptsize $\ln(a_3^2e^{\phi})''-a_1^4a_4^2e^{2\phi}-{\phi}''=0,$}
				&\multirow{2}{11em}{\scriptsize $a_3^2e^{\phi}=\frac{2p_5^2\:e^{-N\tau}}{A^2},$}
                &
				\\

				&\multirow{2}{9em}{\scriptsize $V_4=-\frac{A^2}{2a_1^2a_2^2a_4^2}.$}
				&\multirow{2}{9em}{\scriptsize $\ln(a_4^2 e^{\phi})''=0.$}
				&\multirow{2}{11em}{\scriptsize $a_4^2e^{\phi}=p_4\:e{\frac{N\tau}{2}}.$}
                &
				\\
				
				&
				&
				&
				&
				\\

				\hline

				\multirow{5}{3em}{\scriptsize $ VI_{0}\oplus R$}

				&\multirow{2}{12em}{\scriptsize $V_1=\frac{a_1^2}{2a_2^2a_3^2}-\frac{a_2^2}{2a_1^2a_3^2}-\frac{A^2}{2a_1^2a_2^2a_4^2},$}
				&\multirow{2}{14em}{\scriptsize $\ln(a_1^2 e^{\phi})''+(a_1^4-a_2^4)a_4^2e^{2\phi}=0,$}
				&
\multirow{2}{10em}{\scriptsize $a_1^2 e^{\phi}=a_2^2e^{\phi}=p_1\:e^{N_1\tau},$}

                &\multirow{4}{10em}{\scriptsize $-3N_1^2-2N_1N_4-N_4^2=0.$}
				\\

				 &\multirow{1}{12em}{\scriptsize$V_2=-\frac{a_1^2}{2a_2^2a_3^2}+\frac{a_2^2}{2a_1^2a_3^2}-\frac{A^2}{2a_1^2a_2^2a_4^2}, $}
				&\multirow{2}{18em}{\scriptsize $\ln(a_2^2e^{\phi})''-(a_1^4-a_2^4)a_4^2e^{2\phi}=0,$}
				&\multirow{1}{11em}{\scriptsize $a_3^2e^{\phi}=\frac{2p_3^2\:e^{-N\tau}}{A^2\sinh^4(p_3\tau)},$}
				&
				\\

				&\multirow{2}{11em}{\scriptsize $V_3=-\frac{a_1^2}{2a_2^2a_3^2}-\frac{a_2^2}{2a_1^2a_3^2}-\frac{1}{a_3^2},$}
				&\multirow{3}{18em}{\scriptsize $\ln(a_3^2e^{\phi})''-(a_1^4+a_2^4)a_4^2e^{2\phi}-2a_1^2a_2^2a_4^2e^{2\phi}-\phi''=0,$}
				&\multirow{2}{10em}{\scriptsize $ a_4^2e^{\phi}=p_4\:e^{N_4\tau},$}
				&
				\\

				&\multirow{2}{10em}{\scriptsize $ V_4=-\frac{A^2}{2a_1^2a_2^2a_4^2}.$}
				&\multirow{3}{18em}{\scriptsize $\ln(a_4^2 e^{\phi})''=0.$}
				&

				&
				\\
				
				&
				&
				&
                &
				\\

				\hline

				\multirow{4}{3em}{\scriptsize	$VII_0\bigoplus R$}
				
&\multirow{2}{12em}{\scriptsize $V_1=\frac{a_1^2}{2a_2^2a_3^2}-\frac{a_2^2}{2a_1^2a_3^2}-\frac{A^2}{2a_1^2a_2^2a_3^2},$}
				&\multirow{2}{13em}{\scriptsize $\ln(a_1^2 e^{\phi})''+(a_1^4-a_2^4)a_4^2e^{2\phi}=0,$}
                &\multirow{2}{10em}{\scriptsize $a_1^2e^{\phi}=a_2^2e^{\phi}=L_1^2e^{\frac{p_1\tau}{2}},$}
                &\multirow{4}{10em}{\scriptsize $\frac{p_1^2}{4}+(2N+p_4)p_1+N^2+2p_4N-p_3^2=0.$}
				\\

				&
\multirow{2}{12em}{\scriptsize $V_2=-\frac{a_1^2}{2a_2^2a_3^2}-\frac{a_2^2}{2a_1^2a_3^2}-\frac{A^2}{2a_1^2a_2^2a_3^2},$}
				&\multirow{2}{12em}{\scriptsize $\ln(a_2^2e^{\phi})''-(a_1^4-a_2^4)a_4^2e^{2\phi}=0,$}
				&\scriptsize $a_3^2e^{\phi}=\frac{p_3\:e^{N\tau}}{\cosh(p_3\tau)},$
				&
				\\

				&\multirow{3}{14em}{\scriptsize $V_3=-\frac{a_1^2}{2a_2^2a_3^2}-\frac{a_2^2}{2a_1^2a_3^2}+\frac{1}{a_3^2},$}
				&\multirow{3}{17em}{\scriptsize $\ln(a_3^2e^{\phi})''-(a_1^4+a_2^4)a_4^2e^{2\phi}+2a_1^2a_2^2a_4^2e^{2\phi}-\phi''=0,$}
				&\scriptsize $a_4^2e^{\phi}=L_4\:e^{\frac{N\tau}{2}}.$
				&
				\\

				&\multirow{4}{10em}{\scriptsize $V_4=-\frac{A^2}{2a_1^2a_2^2a_3^2}.$}
				&\multirow{4}{12em}{\scriptsize $\ln(a_4^2 e^{\phi})''=0.$}
				&
				&
				\\

				&
				&
				&
				&

				\\
                &
				&
				&
				&

				\\

				\hline
			\end{tabular}
	
			\caption*{\scriptsize Here for all Lie algebra of this table $H_{124}=A $  and  $\phi=\ln(\frac{p_4}{A^2})-N\tau-\ln(\cosh(p_4\tau)).$}
		\end{threeparttable}
		
	\end{sidewaystable}

	\newpage

	\begin{sidewaystable}[H]
		
		\begin{threeparttable}
			 \caption*{$\text {Table 2}:\hspace{1mm}\textbf{Case}\hspace{1mm}\textbf{ii}:$ continued }
			\centering
			\begin{tabular}{| p{3em} | p{14em} | p{12em} | p{15em} | p{4.5em}  |}
\hline
                \multirow{2}{3em}{\scriptsize \tiny  Lie Algebra  }
				&\multirow{2}{5em} {\scriptsize  potentials }
                &\multirow{2}{6em}{\scriptsize $H_{\mu\nu\lambda}$}
				&\multirow{2}{5.3em} {\scriptsize equations }
				&\multirow{2}{9em} {\scriptsize  solution }
				
				\\
				
				&
                &
			    &
				&
				
				\\
\hline

				\multirow{4}{3em}{\scriptsize $A_{4,5}^{a,b}$}

				&\multirow{2}{11em}{\scriptsize $V_1=-\frac{a+1}{a_4^2}-\frac{A^2}{2a_1^2a_2^2a_4^2},$}
                &\multirow{2}{7em}{\scriptsize $H_{124}=A\:e^{(1+a)\:x_4},$}
				&\multirow{2}{16em}{\scriptsize $\ln(a_1^2 e^{\phi})''-2\left(a+1\right)a_1^2a_2^2a_3^2\:e^{2\phi}=0,$}
				&\multirow{4}{4.5em}{\scriptsize We could not find a solution}
				
				\\

				&\multirow{1}{12em}{\scriptsize $V_2=-\frac{a\left(a+1\right)}{a_4^2}-\frac{A^2}{2a_1^2a_2^2a_4^2},$}
                &\multirow{2}{10em}{\scriptsize $dH=0$ with condition  $b=0,$}
				&\multirow{2}{16em}{\scriptsize $\ln(a_2^2e^{\phi})''-2a\left(a+1\right)a_1^2a_2^2a_3^2\:e^{2\phi}=0,$}
				&
				
				\\
				
				&\multirow{1}{11em}{\scriptsize $V_3=0,$}
                &
				&\multirow{2}{16em}{\scriptsize $\ln(a_3^2e^{\phi})''-\phi''=0,$}
				&
				
				\\
				
				&\multirow{2}{11em}{\scriptsize$V_4=-\frac{a_2^2+1}{a_4^2}-\frac{A^2}{2a_1^2a_2^2a_4^2}.$}
                &
				&\multirow{3}{16em}{\scriptsize $\ln(a_4^2e^{\phi})''-2\left(a^2+1\right)a_1^2a_2^2a_3^2\:e^{2\phi}=0.$}
				&
				
				\\

                &
                &
				&
				&
				
				\\

				\hline
				
				\multirow{4}{3em}{\scriptsize $A_{4,5}^{a,a}$}

				&\multirow{2}{9em}{\scriptsize $V_1=-\frac{1}{a_4^2}-\frac{A^2}{2a_1^2a_2^2a_4^2},$}
                &\multirow{2}{9em}{\scriptsize $H_{134}=A\:e^{(1+a)\:x_4},$}
				&\multirow{2}{14em}{\scriptsize $\ln(a_1^2 e^{\phi})''-2a_1^2a_2^2a_3^2\:e^{2\phi}=0,$}
				&\multirow{4}{4.5em}{\scriptsize We could not find a solution}
				
				\\

				&\multirow{2}{10em}{\scriptsize $V_2=-\frac{A^2}{2a_1^2a_2^2a_4^2},$}
                &\multirow{2}{12em}{\scriptsize $dH=0$  with condition \hspace{2mm} $ a=0,$}
				&\multirow{1}{14em}{\scriptsize $\ln(a_2^2e^{\phi})''=0,$}
				&
				
				\\

				&\multirow{2}{12em}{\scriptsize $V_3=0,$}
                &
				&\multirow{1}{12em}{\scriptsize $\ln(a_3^2e^{\phi})''-\phi''=0,$}
				&
				
				\\

				&\multirow{2}{12em}{\scriptsize $V_4=-\frac{1}{a_4^2}-\frac{A^2}{2a_1^2a_2^2a_4^2}.$}
                &
				&\multirow{1}{12em}{\scriptsize $\ln(a_4^2e^{\phi})''-2a_1^2a_2^2a_3^2\:e^{2\phi}=0.$}
				&
				
				\\
				
				&
                &
				& \multirow{1}{16em}{\scriptsize  eq (8) impose the restriction: $\hspace{2mm} a_1a_2^{a}=a_4^{a+1}$}
				&
				
				\\

                &
                &
				&
				&
				
				\\

				\hline

				\multirow{6}{3em}{\scriptsize $A_{4,9}^{b}$}
				
                &\multirow{2}{10em}{\scriptsize $V_1=\frac{a_1^2}{2a_2^2a_3^2}-\frac{2}{a_4^2}-\frac{A^2}{2a_1^2a_2^2a_4^2},$}
                &\multirow{2}{10em}{\scriptsize $ H_{124}=A\:e^{(2+b)\:x_4},$}
				&\multirow{2}{16em}{\scriptsize$\ln(a_1^2 e^{\phi})''+\left(a_1^4a_4^2-4a_1^2a_2^2a_3^2\right)e^{2\phi}=0,$}
				&\multirow{4}{5em}{\scriptsize We could not find a solution}
				
				\\

				&\multirow{2}{12em}{\scriptsize $V_2=-\frac{a_1^2}{2a_2^2a_3^2}-\frac{2}{a_4^2}-\frac{A^2}{2a_1^2a_2^2a_4^2},$}
                &\multirow{2}{12em}{\scriptsize $dH=0$  with condition \hspace{2mm} $ b=0,$}
				&\multirow{1}{16em}{\scriptsize $\ln(a_2^2e^{\phi})''-\left(a_1^4a_4^2+4a_1^2a_2^2a_3^2\right)e^{2\phi}=0,$}
				&
				
				\\

                &\multirow{3}{10em}{\scriptsize $V_3=-\frac{a_1^2}{2a_2^2a_3^2},$}
                &
				&\multirow{1}{12em}{\scriptsize$\ln(a_3^2e^{\phi})''-a_1^4a_4^2e^{2\phi}-\phi''=0,$}
				&
				
				\\

				&\multirow{4}{12em}{\scriptsize $V_4=-\frac{2}{a_4^2}-\frac{A^2}{2a_1^2a_2^2a_4^2}.$}
                &
				&\multirow{1}{12em}{\scriptsize $\ln(a_4^2e^{\phi})''-4a_1^2a_2^2a_3^2e^{2\phi}=0.$}
				&
				
				\\

				&
                &
				&\multirow{2}{14em}{\scriptsize eq (8) impose the restriction: $\hspace{2mm} a_1a_2=Ba_4^{2}$}
				&
				
				\\

                &
                &
				&
				&
				
				\\

				\hline

				\multirow{6}{3em}{\scriptsize $A_{4,9}^{0}$}
				
                &\multirow{2}{12em}{\scriptsize $V_1=\frac{a_1^2}{2a_2^2a_3^2}-\frac{2}{a_4^2}-\frac{A^2}{2a_1^2a_2^2a_4^2},$}
                &\multirow{2}{7em}{\scriptsize $H_{124}=A\:e^{2x_4},$}
				&\multirow{2}{16em}{\scriptsize$\ln(a_1^2 e^{\phi})''+\left(a_1^4a_4^2-4a_1^2a_2^2a_3^2\right)e^{2\phi}=0,$}
				&\multirow{4}{5em}{\scriptsize We could not find a solution}
				
				\\

				&\multirow{1}{12em}{\scriptsize $V_2=-\frac{a_1^2}{2a_2^2a_3^2}-\frac{2}{a_4^2}-\frac{A^2}{2a_1^2a_2^2a_4^2},$}
                &
				&\multirow{1}{14em}{\scriptsize$\ln(a_2^2e^{\phi})''-\left(a_1^4a_4^2+4a_1^2a_2^2a_3^2\right)e^{2\phi}=0,$}
				&

				\\

                &\multirow{2}{10em}{\scriptsize $V_3=-\frac{a_1^2}{2a_2^2a_3^2},$}
                &
				&\multirow{1}{12em}{\scriptsize$\ln(a_3^2e^{\phi})''-a_1^4a_4^2e^{2\phi}-\phi''=0 ,$}
				&
				
				\\

				&\multirow{3}{12em}{\scriptsize $V_4=-\frac{2}{a_4^2}-\frac{A^2}{2a_1^2a_2^2a_4^2}.$}
                &
				&\multirow{1}{12em}{\scriptsize $\ln(a_4^2 e^{\phi})''-4a_1^2a_2^2a_3^2e^{2\phi}=0.$}
				&
				
				\\

				&
                &
				&\multirow{2}{15em}{\scriptsize eq (8) impose the  restriction:$\hspace{2mm}a_1a_2=Ba_4^{2}$}
				&
				
				\\

                &
                &
				&
				&
				
				\\

				\hline
			\end{tabular}
\caption*{\scriptsize Here for all Lie algebra of this table $\phi=\ln(\frac{p_4}{A^2})-N\tau-\ln(\cosh(p_4\tau)).$}
\label{table:4}
		\end{threeparttable}
		
	\end{sidewaystable}
\newpage

	\begin{sidewaystable}[H]
		
		\begin{threeparttable}
			 \caption{$\hspace{-1mm}\textbf{Case}\hspace{1mm}
\textbf{iii}:\hspace{2mm}H=A\hspace{1mm}
(g^{-1}dg)^1\wedge(g^{-1}dg)^3\wedge(g^{-1}dg)^4,\:\:
\hspace{3mm}{}^{\star}db=e^{2\phi}dx^{2}\wedge dt,$}
			\centering
			\begin{tabular}{| p{3em} | p{8.8em}| p{7.5em} |  p{12.5em} | p{9em} | p{8.5em} |}
\hline
 \multirow{2}{5.3em}{\scriptsize\tiny  Lie Algebra  }
				&\multirow{2}{5em} {\scriptsize  potentials }
                &\multirow{2}{6em}{\scriptsize $H_{\mu\nu\lambda}$}
				&\multirow{2}{5.3em} {\scriptsize equations }
				&\multirow{2}{9em} {\scriptsize  solution }
				&\multirow{2}{6em}{\scriptsize condition from eq (16)}
				\\
				
				&
                &
			    &
				&
				&
				\\
\hline

				\multirow{4}{3em}{\scriptsize $I\oplus R$}

				&\multirow{2}{10em}{\scriptsize $V_1=V_3=V_4=-\frac{A^2}{2a_1^2a_3^2a_4^2},$}
                &\multirow{2}{7.5em}{\scriptsize $H_{134}=A,$}
				&\multirow{2}{10em}{\scriptsize $\ln(a_2^2 e^{\phi})''-\phi''=0,$}
				&\multirow{2}{11em}{\scriptsize $a_2^2e^{\phi}=\frac{p_4e^{N_4\tau}}{\cosh(p_4\tau)},$}
				&\multirow{4}{8.5em}{\scriptsize $ N_4^2+2(N_1+N_2+N_3)N_4+(N_2+N_3)N_1+N_2N_3-p_4^2=0.$}
				\\

				&\multirow{1}{12em}{\scriptsize $V_2=0,$}
                &
				&\scriptsize $\ln(a_c^2 e^{\phi})''=0,$
				&\scriptsize $a_c^2e^{\phi}=p_ce^{N_c\tau},$
				&
				\\
				
				&
                &
				&\scriptsize $\c=1,3,4.$
				&\scriptsize $\c=1,3,4.$
				&
				\\
				
				&
                &
				&
				&
				&
				\\

				\hline
				
				\multirow{4}{3em}{\scriptsize $A_{4,5}^{a,b}$}

				&\multirow{2}{9em}{\scriptsize $V_1=-\frac{b+1}{a_4^2}-\frac{A^2}{2a_1^2a_3^2a_4^2},$}
                &\multirow{2}{7.5em}{\scriptsize $H_{134}=A\:e^{(1+b)\:x_4},$}
				&\multirow{2}{13em}{\scriptsize $\ln(a_1^2 e^{\phi})''-2\left(b+1\right)a_1^2a_2^2a_3^2\:e^{2\phi}=0,$}
				&\multirow{2}{10em}{\scriptsize $a_1^2e^{\phi}=p_1^2\cosh^{b_1}(p_3\tau)e^{N_1\tau},$}
				&\multirow{2}{9em}{\scriptsize $b_4(b+1)+2b+3b_1=0,$}
				\\

				&\multirow{1}{10em}{\scriptsize $V_2=0,$}
                &\multirow{2}{7.5em}{\scriptsize $dH=0$ \tiny with condition $ b=0$}
				&\multirow{1}{13em}{\scriptsize $\ln(a_2^2e^{\phi})''-\phi''=0,$}
				&\multirow{1}{10em}{\scriptsize $a_2^2e^{\phi}=\frac{p_2e^{N_2\tau}}{\cosh(p_2\tau)},$}
				&\multirow{2}{8.5em}{\scriptsize $ -4A^2p_3^2p_1^2(b^2+b+1)-(b_4+b_3)b_1p_2^2-b_4b_3p_2^2=0.$}
				\\

				&\multirow{1}{10em}{\scriptsize $V_3=-\frac{b\left(b+1\right)}{a_4^2}-\frac{A^2}{2a_1^2a_3^2a_4^2},$}
                &
				&\multirow{1}{13em}{\scriptsize $\ln(a_3^2e^{\phi})''-2b\left(b+1\right)a_1^2a_2^2a_3^2\:e^{2\phi}=0,$}
				&\multirow{2}{12em}{\scriptsize $a_3^2e^{\phi}=p_3^2\cosh^{b_3}(p_2\tau)e^{N_3\tau},$}
				&
				\\

				&\multirow{2}{10em}{\scriptsize $V_4=-\frac{b_2^2+1}{a_4^2}-\frac{A^2}{2a_1^2a_3^2a_4^2}.$}
                &
				&\multirow{1}{13em}{\scriptsize $\ln(a_4^2e^{\phi})''-2\left(b^2+1\right)a_1^2a_2^2a_3^2\:e^{2\phi}=0.$}
				&\multirow{2}{10em}{\scriptsize $ a_4^2e^{\phi}=p_4^2\cosh^{b_4}(p_2\tau)e^{N_4\tau}.$}
				&
				\\
				
				&
                &
				&\multirow{2}{13em}{\scriptsize\tiny eq (8) impose the restriction:
 $\hspace{2mm}a_1a_2^{a}=a_4^{a+1}.$}
				&
				&
				\\

                &
                &
				&
				&
				&
				\\

				\hline

				\multirow{6}{3em}{\scriptsize $A_{4,5}^{a,1}$}
				
                &\multirow{2}{10em}{\scriptsize $V_1=V_3=-\frac{2}{a_4^2}-\frac{A^2}{2a_1^2a_3^2a_4^2},$}
                &\multirow{2}{7.5em}{\scriptsize $ H_{134}=A\:e^{2\:x_4}.$}
				&\multirow{2}{12em}{\scriptsize$\ln(a_1^2 e^{\phi})''-4a_1^2a_2^2a_3^2\:e^{2\phi}=0,$}
				&\multirow{4}{5em}{\scriptsize We could not find a solution}
				&
				\\

				&\multirow{1}{12em}{\scriptsize $V_2=0,$}
                &\multirow{2}{7.5em}{\scriptsize $dH=0$  with condition  $a=0$}
				&\multirow{1}{16em}{\scriptsize $\ln(a_2^2e^{\phi})''-\phi''=0,$}
				&
				&
				\\

                &\multirow{1}{10em}{\scriptsize $V_4=-\frac{2}{a_4^2}-\frac{A^2}{2a_1^2a_3^2a_4^2}.$}
                &
				&\multirow{1}{12em}{\scriptsize $ \ln(a_3^2e^{\phi})''-4a_1^2a_2^2a_3^2\:e^{2\phi}=0,$}
				&
				&
				\\

				&
                &
				&\multirow{1}{12em}{\scriptsize $\ln(a_4^2e^{\phi})''-4a_1^2a_2^2a_3^2\:e^{2\phi}=0.$}
				&
				&
				\\

				&
                &
				&\multirow{1}{14em}{\scriptsize eq (8) impose the  restriction: $\hspace{2mm}a_1a_3=a_4^{2}.$}
				&
				&
				\\

                &
                &
				&
				&
				&
				\\

				\hline
			\end{tabular}
\caption*{\scriptsize Here for all Lie algebra of this table $\phi=\ln(\frac{p_2}{A^2})-N_2t-\ln(\cosh(p_2t)).$}
\label{table:4}
		\end{threeparttable}
		
	\end{sidewaystable}

	\newpage

	\begin{sidewaystable}[H]
		
		\begin{threeparttable}
			 \caption{$\hspace{-1mm}\textbf{Case}\hspace{1mm}
\textbf{iv}:\hspace{2mm}H=A\hspace{1mm}
(g^{-1}dg)^2\wedge(g^{-1}dg)^3\wedge(g^{-1}dg)^4,\:\:
\hspace{3mm}{}^{\star}db=e^{2\phi}dx^{1}\wedge dt,$}
			\centering
			\begin{tabular}{| p{2em} | p{13.5em} | p{3em} | p{17em} | p{9em} | p{4.5em} |}
\hline
                \multirow{2}{2em}{\scriptsize\tiny  Lie Algebra  }
				&\multirow{2}{5em} {\scriptsize  potentials }
                &\multirow{2}{3em}{\scriptsize $H_{\mu\nu\lambda}$}
				&\multirow{2}{5.3em} {\scriptsize equations }
				&\multirow{2}{9em} {\scriptsize  solution }
				&\multirow{2}{4.5em} {\scriptsize
\tiny condition from eq (16)}
				\\
				
				&
                &
                &
			    &
				&
				
				\\

\hline

				\multirow{4}{2em}{\scriptsize\tiny $I\oplus R$}

				&\multirow{4}{15em}{\scriptsize $V_2=V_3=V_4=-\frac{A^2}{2a_2^2a_3^2a_4^2},$}
                &\multirow{2}{3em}{\scriptsize $H_{234}=A,$}
				&\multirow{2}{15em}{\scriptsize $\ln(a_1^2 e^{\phi})''-\phi''=0,$}
				&\multirow{2}{11em}{\scriptsize $a_1^2e^{\phi}=\frac{p_4e^{N_1\tau}}{\cosh(p_4\tau)},$}
				&\multirow{4}{4.5em}{\scriptsize\tiny $ N_1^2+2(N_1+N_2+N_3)N_1+(N_2+N_3)N_1+N_2N_3-p_4^2=0.$}
				\\

				&\multirow{6}{15em}{\scriptsize $V_1=0,$}
                &
				&\scriptsize $\ln(a_c^2 e^{\phi})''=0,$
				&\scriptsize $a_c^2e^{\phi}=p_ce^{N_c\tau},$
				&
				\\
				
				&
                &
				&\scriptsize $ c= 2,3,4.$
				&\scriptsize $ c= 2,3,4.$
				&
				\\
				
				&
                &
				&
				&
				&
				\\

                &
                &
				&
				&
				&
				\\

				\hline
				
				\multirow{4}{2em}{\scriptsize\tiny $A_{4,6}^{a,b}$}

				&\multirow{2}{15em}{\scriptsize $V_1=0,$}
                &
				&\multirow{2}{15em}{\scriptsize $\ln(a_1^2 e^{\phi})''-\phi''=0,$}
				&\multirow{2}{12em}{\scriptsize $a_2^2e^{\phi}=a_3^2e^{\phi}=a_4^2e^{\phi}=B^2,$}
				&\multirow{2}{4.5em}{\scriptsize $N^2=P_4^2.$}
				\\

				&\multirow{3}{15em}{\scriptsize $V_2=\frac{a_2^2}{2a_3^2a_4^2}-\frac{a_3^2}{2a_2^2a_4^2}-\frac{2b^2}{a_4^2}-\frac{A^2}{2a_2^2a_3^2a_4^2},$}
                &\multirow{1}{3em}{\scriptsize\tiny $ H_{234}=A\:e^{2b\:x_4}$}
				&\multirow{1}{15em}{\scriptsize $\ln(a_2^2e^{\phi})''-a_1^2\left(a_3^4-a_2^4+4b^3a_2^2a_3^2\right)e^{2\phi}=0,$}
				&\multirow{1}{10em}{\scriptsize $a_1^2e^{\phi}=\frac{e^{N\tau}}{\cosh(p_4\tau)}.$}
				&
				\\

				&\multirow{5}{15em}{\scriptsize $V_3=-\frac{a_2^2}{2a_3^2a_4^2}+\frac{a_3^2}{2a_2^2a_4^2}-\frac{2b^2}{a_4^2}-\frac{A^2}{2a_2^2a_3^2a_4^2},$}
                &
				&\multirow{2}{15em}{\scriptsize $\ln(a_3^2e^{\phi})''-a_1^2\left(a_2^4-a_3^4+4b^3a_2^2a_3^2\right)e^{2\phi}=0,$}
				&
				&
				\\

				&\multirow{7}{14em}{\scriptsize $V_4=-\frac{a_2^2}{2a_3^2a_4^2}-\frac{a_3^2}{2a_2^2a_4^2}-\frac{(2b^2-1)}{a_4^2}-\frac{A^2}{2a_2^2a_3^2a_4^2}.$}
                &
				&\multirow{2}{18em}{\scriptsize $\ln(a_4^2e^{\phi})''-a_1^2\left(a_2^4+a_3^4+2(2b^2-1)a_2^2a_3^2\right)e^{2\phi}=0.$}
				&
				&
				\\
				
				&
                &
				&\multirow{3}{14em}{\scriptsize eq (2b) led to the  condition  \hspace{7mm} $a=0 $,}
				&
				&
				\\

                &
                &
				&\multirow{4}{14em}{\scriptsize eq (8) impose the restriction: $\hspace{10mm}(a_2a_3)^{b}=Ba_4^{2b},$}
				&
				&
				\\
				
                &
                &
                &
			    &
				&
				
				\\

                &
                &
                &
			    &
				&
				
				\\

				\hline
			
			\end{tabular}
\caption*{\scriptsize Here for all Lie algebra of this table $\phi=-N\tau-\ln(\cosh(p_4\tau)).$}
\label{table:4}
		\end{threeparttable}
		
	\end{sidewaystable}

	\newpage

\vspace{10mm}

$\bullet$ For the Lie algebra $A_{4,2}^{1}$  the constraints imposed by eqs (8)and (9) have the following forms:

$ a_1a_2a_3=D a_4^{3}$ and \hspace{1mm} $a_2=0,$  where $D$ is a constant. No solution exists for any of the cases i-iv for this Lie algebra

\vspace{5mm}
$\bullet$ For the Lie algebra $A_{4,3}$  the constraints imposed by eqs (8)and (9) have the following  forms:

$ a_1=D a_4$ and  \hspace{1mm} $a_2=0,$  where $D$ is a constant. No solution exists for any of the cases i-iv for this Lie algebra

\vspace{5mm}
$\bullet$ For the Lie algebra $A_{4,4}$  the constraints imposed by eqs (8)and (9) have the following forms:

$ a_1a_2a_3=D a_4^3$ and  \hspace{1mm} $a_1=0,$ where $D$ is a constant. No solution exists for any of the cases i-iv for this Lie algebra

\vspace{5mm}
$\bullet$ For the Lie algebra $A_{4,5}^{(a,b)}$  for the  case iv eq (2b) leads to a contradiction.

\vspace{5mm}
$\bullet$ For the Lie algebras, $A_{4,5}^{(a,a)}$,\hspace{1mm} $II\oplus R$,\hspace{1mm} $VI_{0}\oplus R$,\hspace{1mm}$VII_{0}\oplus R$ and $A_{4,9}^{b}$  for  cases iii and iv eq (2b) leads to a contradiction.

\vspace{5mm}
$\bullet$ For the Lie algebra $A_{4,5}^{(a,1)}$  for  cases ii and iv eq (2b) leads to a contradiction.

\vspace{5mm}
$\bullet$ For the Lie algebras $A_{4,5}^{(1,1)}$  ,$ A_{4,9}^1$  and $A_{4,12}$  for all cases i-iv and Lie algebras  $ A_{4,9}^0$  for  cases i,iii and iv the results from eq (2b) is $A=0$ which is a contradiction.

\vspace{5mm}
$\bullet$ For the Lie algebra $A_{4,10}$,\hspace{1mm}$A_{4,8}$,\hspace{1mm}$A_{4,11}^{a},$ and $II\oplus R$  for cases ii,iii and iv eq (2b) leads to a contradiction.

\vspace{5mm}
$\bullet$For the Lie algebras $A_{4,12}$,\hspace{1mm}$A_2\oplus A_2$,\hspace{1mm}$V \oplus R $,\hspace{1mm}$IX\oplus R$,\hspace{1mm} $VIII\oplus R$ and $A_{4,1}$  for all cases i-iv eq (2b) leads to a contradiction.

\vspace{5mm}
$\bullet$ For the Lie algebra $A_{4,6}^{(a,b)}$  for  cases ii and iii eq (2b) leads to a contradiction.

\vspace{5mm}
$\bullet$ For the Lie algebra $A_{4,7}$  the constraints imposed by eqs (14)and (15) have the following forms:

$ a_1^2a_2a_3=B a_4^4$, and $a_2=0,$  where $B$ is a constant. No solution exists for any of the cases i-iv for this Lie algebra

\vspace{5mm}
$\bullet$  For the Lie algebra $III\oplus R$  the constraints imposed by eqs (8)and (9) have the following form: $a_2a_3=D a_1^2$
where $D$ is a constant. No solution exists for any of the cases i-iv for this Lie algebra
\vspace{5mm}
$\bullet$ For the Lie algebra $IV \oplus R$   eqs (8)and (9) impose constraints  $a_2a_3=D a_1^2$ and $a_3=0,$  where $D$ is a constant.
No solution exists for any of the cases i-iv for this Lie algebra

\vspace{5mm}
$\bullet$ For the Lie algebra $VI_{b} \oplus R$  the constraints imposed by eqs (8)and (9) have the following forms:
$ a_2a_3=D a_1^2$ and $ a_2^2+a_3^2=0,$ where $D$ is a constant. No solution exists for any of the cases i-iv for this Lie algebra

\vspace{5mm}
$\bullet$ For the Lie algebra $A_{4,2}^{b}$  the constraints imposed by eqs (8)and (9) have the following  forms:
$ a_1^ba_2a_3=B a_4^{b+2}$ and $ a_2=0,$  where $B$ is a constant. No solution exists for any of the cases i-iv for this Lie algebra
\\
\section{ Example  $VII_0 \bigoplus R$ as a 4+1 dimensional cosmology coupled to dilaton and antisymmetric matter}
 $\; \; $ In this section, following the cosmological model proposed in \cite{Mojaveri}, we consider the model with Lie group ($VII_0 \bigoplus R$) for the \textbf{Case}\hspace{1mm} \textbf{i} discussed in the previous section {\footnote{The reason for considering this model, in addition of campactification results in extra dimension, is that it is the simplest model (in terms of calculation) with respect to the other models obtained from other real four dimensional Lie groups.}}. It can be demonstrated that in D dimension  the equations $(2a-2c)$ can be expressed as following Einstein equations\cite{Frad,Ca,Fa,Naderi}:
\renewcommand\theequation{\arabic{tempeq}\alph{equation}}
\setcounter{equation}{-1} \addtocounter{tempeq}{1}
\begin{eqnarray}
&&\hspace{-100mm}\widetilde{R}_{\mu\nu}-\frac{1}{2}\widetilde{R}\widetilde{g}_{\mu\nu}=
\kappa_D^2({T_{\mu\nu}}^{(\phi)}+T_{\mu\nu}^{(H)}),
\end{eqnarray}
where the metric $\widetilde{g}$ and $g$ are related as\cite{Fa}:
\renewcommand\theequation{\arabic{tempeq}\alph{equation}}
\setcounter{equation}{-1} \addtocounter{tempeq}{1}
\begin{eqnarray}
&&\hspace{-140mm}\widetilde{g}_{\mu\nu}=e^\frac{2{\phi}}{D-2} g_{\mu\nu}.
\end{eqnarray}
such that in Einstein frame the effective action for the bosonic string in D dimension can be written as \cite{Frad,Ca}:
\renewcommand\theequation{\arabic{tempeq}\alph{equation}}
\setcounter{equation}{-1} \addtocounter{tempeq}{1}
\begin{eqnarray}
S=\frac{-1}{2\kappa_D^2}\int d^{D}x \sqrt{\widetilde{g}}(\widetilde{R}-\frac{(\widetilde{\nabla}
\phi)^2}{D-2}-\frac{1}{12}e^\frac{4{\phi}}{D-2}H^2)
\end{eqnarray}
and the energy- momentum tensor has two part, first
 part is related to the  dilaton field $\phi$ and the second part
is constructed from torsion of anti-symmetric matter field $B$ as follows\cite{Frad,Ca,Naderi}:
\renewcommand\theequation{\arabic{tempeq}\alph{equation}}
\setcounter{equation}{-1} \addtocounter{tempeq}{1}
\begin{eqnarray}
&&\hspace{-80mm}\kappa_D^2{T_{\mu\nu}}^{(\phi)}=\frac{1}{D-2}(\widetilde{\nabla}_{\mu}\phi \widetilde{\nabla}_{\nu}\phi-\frac{1}{2}(\widetilde{\nabla}\phi)^2\widetilde{g}_{\mu\nu}),\nonumber\\
&&\hspace{-80mm}\kappa_D^2
T_{\mu\nu}^{(H)}=\frac{e^\frac{4{\phi}}{D-2}}{4}(H_{\mu\kappa\lambda}H_{\nu}^{\kappa\lambda}-\frac{1}{6}H^2\widetilde{g}_{\mu\nu}).
\end{eqnarray}

Using $(11)$ and $(13)$ with assuming the condition:
\begin{eqnarray}
p_1+p_3+3N+5p_4=0,
\end{eqnarray}
one can obtain the relation between $\tau$ and $t$ as follows:
\begin{eqnarray}
\tau=-\frac{1}{2p_4}\ln(C t^\frac{2}{3}-1),
\end{eqnarray}
where $C=(\frac{-18{p_4}^3}{L_1 L_2A^4})^{\frac{1}{3}}$ and we must have $p_4<0$. Then, one can express all scale factors $a_i$ of table 1 \textbf{Case}\hspace{1mm} \textbf{i} for $VII_0\bigoplus R$ in term of $t$, as follows:
\renewcommand\theequation{\arabic{tempeq}\alph{equation}}
\setcounter{equation}{-1} \addtocounter{tempeq}{1}
\begin{eqnarray}
&\hspace{-80mm}a_1=a_2=\frac{AL_1^\frac{1}{4}\sqrt{C}}{\sqrt{-2 p_4}}(Ct^\frac{2}{3}-1)^{-\frac{p_1+2N+2p_4}{8p_4}}t^\frac{1}{3},\nonumber\\
&\hspace{-90mm}a_3=\frac{A\sqrt{C L_2}}{\sqrt{-2 p_4}}(Ct^\frac{2}{3}-1)^{-\frac{p_3+N+p_4}{4p_4}}t^\frac{1}{3},\nonumber\\
&\hspace{-115mm}a_4=A(Ct^\frac{2}{3}-1)^{-\frac{N}{2p_4}},\nonumber\\
&\hspace{-80mm}\phi=\ln(\frac{-p_4}{A^2})-\ln(\frac{Ct^\frac{2}{3}}{2(Ct^\frac{2}{3}-1)^\frac{1}{2}})+\frac{N}{2p_4}\ln(Ct^\frac{2}{3}-1).
\end{eqnarray}
also the components of energy-momentum tensor $(20)(20a)(20b)$ (in D=5) take the following forms:
\renewcommand\theequation{\arabic{tempeq}\alph{equation}}
\setcounter{equation}{-1} \addtocounter{tempeq}{1}
\begin{eqnarray}
&\hspace{-115mm}\kappa^2T_{00}^\phi+\kappa^2T_{00}^H=\frac{1}{54p_4^2 t^2}(2-(\frac{-2p_4}{A^2Ct^\frac{2}{3}})^\frac{2}{3}(Ct^\frac{2}{3}-1)^{\frac{N+p_4}{3p_4}})(\frac{p_4(2-Ct^\frac{2}{3})+CNt^\frac{2}{3}}{Ct^\frac{2}{3}-1})^2
+\frac{-8p_4^{5}(Ct^\frac{2}{3}-1)^\frac{5(N+p_4)+p_1+p_3}{2p_4}}{A^8L_1 L_2 (Ct^\frac{2}{3})^5},\\
&\hspace{-130mm}\kappa^2T_{11}^\phi+\kappa^2T_{11}^H=\kappa^2T_{22}^\phi+\kappa^2T_{22}^H=\frac{(2A)^\frac{2}{3}\sqrt{L_1}(Ct^\frac{2}{3}-1)^{-\frac{3p_1+2(N+p_4)}{12p_4}-2}}{108(-p_4)^\frac{7}{3}t^2}(p_4(2-Ct^\frac{2}{3})+CNt^\frac{2}{3})^2(ct^\frac{2}{3})^\frac{1}{3}+\nonumber\\
\frac{(16p_4^{10})^\frac{1}{3}}{A^\frac{14}{3}\sqrt{L_1}L_2(Ct^\frac{2}{3})^\frac{10}{3}}(2-(\frac{-2p_4}{A^2Ct^\frac{2}{3}})^\frac{2}{3}(Ct^\frac{2}{3}-1)^{\frac{N+p_4}{3p_4}})(Ct^\frac{2}{3}-1)^\frac{20(N+p_4)+6p_3+3p_1}{12p_4},\\
&\hspace{-160mm}\kappa^2T_{33}^\phi+\kappa^2T_{33}^H=(2A)^\frac{2}{3}\frac{L_2(Ct^\frac{2}{3}-1)^{-\frac{N+p_4+3p_3}{6p_4}-2}}{108(-p_4)^\frac{7}{3}t^2}(p_4(2-Ct^\frac{2}{3})+CNt^\frac{2}{3})^2(Ct^\frac{2}{3})^\frac{1}{3}+\nonumber\\
\frac{2^\frac{4}{3}p_4^\frac{10}{3}}{A^\frac{14}{3}L_1(Ct^\frac{2}{3})^\frac{10}{3}}(2-(\frac{-2p_4}{{A^2}})^\frac{2}{3}\frac{(Ct^\frac{2}{3}-1)^{\frac{N+p_4}{3p_4}}}{(Ct^\frac{2}{3})^\frac{2}{3}})(Ct^\frac{2}{3}-1)^{\frac{3p_1+10(N+p_4)}{6p_4}},\\
&\hspace{-130mm}\kappa^2T_{44}^\phi+\kappa^2T_{44}^H=\frac{1}{54}(\frac{2A}{{p_4}^2})^\frac{2}{3}\frac{(Ct^\frac{2}{3}-1)^{-\frac{2N+5p_4}{3p_4}}}{t^2(Ct^\frac{2}{3})^\frac{2}{3}}(p_4(2-Ct^\frac{2}{3})+CNt^\frac{2}{3})^2+\frac{8(-p_4)^5(Ct^\frac{2}{3}-1)^\frac{3N+\textcolor{red}{5}p_4+p_3+p_1}{2p_4}}{A^6L_1L_2(Ct^\frac{2}{3})^5}.
\end{eqnarray}
Now, by comparing the above energy-momentum tensor with that of a perfect fluid (considered as dark energy)
\renewcommand\theequation{\arabic{tempeq}\alph{equation}}
\setcounter{equation}{-1} \addtocounter{tempeq}{1}
\begin{eqnarray}
T=\left(
    \begin{array}{ccccc}
      -\rho(t) & 0 & 0 & 0 & 0 \\
      0 & P(t) & 0 & 0 & 0 \\
      0 & 0 & P(t) & 0 & 0 \\
      0 & 0 & 0 & P(t) & 0 \\
      0 & 0 & 0 & 0 & P_4(t) \\
    \end{array}
  \right),
\end{eqnarray}
we have $p_1=2p_3$ and $L_2=L_1^\frac{1}{2}$. After employing (20a) and conditions derived from (16) for $VII_0\bigoplus R$ (Table 1, Case i), we obtain $p_3+N=-\frac{5}{3}p_4$ and $p_4=\pm\sqrt{\frac{3}{11}}N$. Therefore, we must consider two cases: $p_4=\pm\sqrt{\frac{3}{11}}N$. It is worth noting that for these two cases, we have:
\begin{equation}
a_1=a_2=a_3=(18A^2t^2(Ct^\frac{2}{3}-1))^{\frac{1}{6}},
\end{equation}
\begin{equation}
\omega(t)=-\frac{1}{2\sqrt2p_4}(18A^7t^2(Ct^\frac{2}{3}-1))^{\frac{1}{3}},
\end{equation}
where $\omega(t)$ is the parameter of the equation of state.
\\
\textbf{Case a :}  $p_4=\sqrt{\frac{3}{11}}N$ so we must have $N<0$; in this case
\begin{equation}
P(t)=\omega(t)\rho(t),
\end{equation}
with
\begin{equation}
P(t)=(\frac{9L_1^3A^2}{4})^\frac{1}{9}
\frac{(Ct^\frac{2}{3}-1)^{\frac{1}{3}(\alpha-5)}}{54t^\frac{16}{9}}(\beta Ct^\frac{2}{3}+2)^2+\\
(\frac{4}{9^{10}A^4})^\frac{1}{9}L_1^\frac{2}{3}t^\frac{-20}{9}(2-(\frac{4p_4^2}{A^4})^\frac{1}{3}\frac{(Ct^\frac{2}{3}-1)^{\frac{\alpha}{2}}}{(Ct^\frac{2}{3})^\frac{2}{3}}(Ct^\frac{2}{3}-1)^\frac{2(\alpha-1)}{3},
\end{equation}
\begin{equation}
\rho(t)=\frac{-1}{54 t^2}(\frac{\beta Ct^\frac{2}{3}+2}{Ct^\frac{2}{3}-1})^2(2-(\frac{4p_4^2}{A^4})^\frac{1}{3}\frac{(Ct^\frac{2}{3}-1)^{\frac{\alpha}{3}}}{(Ct^\frac{2}{3})^\frac{2}{3}})-\\
\frac{8p_4^5}{A^8L_1^{\frac{3}{2}}}\frac{(Ct^\frac{2}{3}-1)^{\alpha-1}}{(Ct^\frac{2}{3})^5},
\end{equation}
\begin{equation}
a_4=A(Ct^\frac{2}{3}-1)^{-\sqrt{\frac{11}{12}}},
\end{equation}
and
\begin{equation}
P_4(t)=F_+(\rho)=(4A^2p_4^2)^\frac{1}{3}\frac{(Ct^\frac{2}{3}-1)^{\frac{-2\alpha}{3}-1}}{54t^2 (Ct^\frac{2}{3})^\frac{2}{3}}(\beta Ct^\frac{2}{3}+2)^2-\frac{8p_4^5}
{A^6L_1^\frac{3}{2}(Ct^\frac{2}{3})^5},
\end{equation}
where $\alpha=\frac{N+p_4}{p_4}=\sqrt{\frac{11}{3}}+1$ and $\beta=\frac{N-p_4}{p_4}=\sqrt{\frac{11}{3}}-1$.
We observe that the parameter of the equation of state is a function of time\cite{RR}. Additionally, $P_4 = F_+(\rho)$ is a function of $\rho$ \cite{NO}. Note that all of these parameters have singularities at $t_{s_1}$ and $t_{s_2}$.
\begin{equation}
t_{s_1}=0 , t_{s_2}=C^{-\frac{2}{3}},
\end{equation}
such that for $t \rightarrow t_{s_1},t_{s_2}$, we have $\omega \rightarrow 0$, $P, P_4 \rightarrow \infty$, $|\rho| \rightarrow \infty$, $a_1, a_2, a_3 \rightarrow 0$ \footnote{Note that because $p_4<0$ we have $e^\phi=\frac{-2p_4}{A^2Ct^\frac{2}{3}}(Ct^\frac{2}{3}-1)^\frac{2}{3}$} , and $a_4 \rightarrow \infty$. In other words, we experience a 'Big Bang' in the $a_i$ space directions and a 'Big Rip' \cite{NO} in the extra space direction. On the other hand, for $t \rightarrow \infty$, we have $a_4 \rightarrow 0$ and $a_i \simeq t^\frac{4}{9}$ i.e. we have compactification for the extra dimension.
Furthermore, the declaration parameters depend on time:
\renewcommand\theequation{\arabic{tempeq}\alph{equation}}
\setcounter{equation}{-1} \addtocounter{tempeq}{1}
\begin{eqnarray}
&\hspace{-80mm}q_i(t)=-\frac{\ddot{a_i(t)}a_i(t)}{\dot{a_i(t)}^2}=\frac{-\frac{5}{3}t^\frac{2}{3}(Ct^\frac{2}{3}-\frac{3}{4})+\frac{3}{2}(Ct^\frac{2}{3}-1)}{\frac{4}{3}(Ct^\frac{2}{3}-\frac{3}{4})^2}
\end{eqnarray}
\renewcommand\theequation{\arabic{tempeq}\alph{equation}}
\setcounter{equation}{-1} \addtocounter{tempeq}{1}
\begin{eqnarray}
q_4(t)=-\frac{\ddot{a_4(t)}a_4(t)}{\dot{a_4(t)}^2}=-\frac{Ct^\frac{2}{3}-1+\frac{2}{3}(\sqrt{\frac{11}{12}}+1)Ct^\frac{-1}{3}}{2\sqrt{\frac{11}{12}}Ct^\frac{2}{3}},
\end{eqnarray}
such that we have $q_i|_{t=t_{s_1}}=-2<0$ and $q_4|_{t=t_{s_1}}=\frac{-1}{3t}(1+\sqrt\frac{12}{11})\rightarrow -\infty$ while $q_i|_{t=t_{s_2}} = -5{t_{s_2}}^\frac{2}{3} < 0$ and $q_4|_{t=t_{s_2}}-\frac{\sqrt{\frac{11}{12}}+1}{3t_s\sqrt{\frac{11}{12}}}< 0$. Furthermore, for $t \rightarrow \infty$, we have $q_i = -\frac{5}{4C} < 0$ and $q_4 = -\sqrt{\frac{3}{11}} < 0$, i.e., we have expansion in all directions which is contradiction with compactification in extra dimension, so this case is not good.

\textbf{Case b :} $p_4=-\sqrt{\frac{3}{11}}N$ with $N>0$. For this case the $P(t)$, $P_4(t)$ and $\rho(t)$ are the same as (23d)-(23g) with replacement $\alpha\leftrightarrow-\beta$ while for $a_4$ we have
\begin{equation}
a_4=A(Ct^\frac{2}{3}-1)^{\sqrt{\frac{11}{12}}},
\end{equation}
such that for $t \rightarrow t_s$ we have $P, P_4\rightarrow \infty$,$|\rho|\rightarrow \infty$, $a_1, a_2, a_3\rightarrow 0$ and $a_4\rightarrow 0$.
i.e we have a $Big Bang$ in all space and extra directions. The declaration parameters $q_i$'s are the same as above while for $q_4$ we have
\begin{eqnarray}
q_4(t)=-\frac{\ddot{a_4(t)}a_4(t)}{\dot{a_4(t)^2}}=-\frac{Ct^\frac{2}{3}-1+\frac{2}{3}(1-\sqrt{\frac{11}{12}})Ct^\frac{-1}{3}}{2\sqrt{\frac{11}{12}}Ct^\frac{2}{3}}
\end{eqnarray}
where ${q_4}{|_{t=t_{s_1}}}=-\frac{\sqrt\frac{12}{11}-1}{3t}\rightarrow-\infty $ we have ${q_4}{|_{t=t_{s_2}}}=\frac{1-\sqrt{\frac{11}{12}}}{3t_{s_2}\sqrt{\frac{11}{12}}}>0$ and furthermore for $t\rightarrow\infty$ we have $q_i = -\frac{5}{4C} < 0$, $q_4=\sqrt{\frac{12}{44}}>0$ i.e. we have expansion in three space directions and contraction in extra dimension which is compatible  with compactification in this direction. Note that all of the physical parameters depend on the magnitude of $B$ field i.e $A$.
\section{conclusion:}
We have obtained  equations of motion of all homogeneous anisotropic string cosmological models characterized by five-dimensional space-time metrics (with real four dimensional isometry Lie groups) and non-vanishing dilaton and anti-symmetric $B$ field, considering different cases of components of their field strength. We have obtained exact solutions for some of these models. For the model with Lie group $VII_0 \oplus R$, we demonstrate that the model can be identified as perfect fluid with time dependent parameter in the equation of state (23b) and (23c), such that the model can be reduced to a $3+1$-dimensional cosmological model, where the extra dimension is compactified; it is shown that all physical parameters depend on the magnitude of $B$ field i.e $A$. One can also construct such model with the Lie group $I\oplus R$ for the $\bf{cases  i,...,iv}$ and with the Lie group $V\oplus R$ for the $\bf{case  i}$, and their relations to FRW model. For the Lie group $A_{4,8}$, one can obtain a solution using the method of Poisson-Lie plurality \cite{Von} on the previous solution obtained for zero $B$ field in \cite{Mojaveri}. Some of these works are under investigation.

\newpage
\vspace{-60mm}{\bf AppendixA:} 4-dimensional real Lie algebras and
their  left invariant one forms\cite{Mojaveri}.
\begin{center}
    \begin{tabular}{l l l l  l l p{15mm} }
    \hline\hline
   \vspace{-1mm}
{\scriptsize ${\bf Algebra}$ }&{\scriptsize Non-zero commutation
relations } & &{\scriptsize $g^{-1} dg$}
\\\hline

\vspace{1mm}

{\scriptsize ${ 4A_1}$}& {\scriptsize $[X_i,X_j]=0$}&
 & {\scriptsize $dx^i X_i$}\\

\vspace{-1mm}

{\scriptsize ${ A_2+2A_1}$}& {\scriptsize$[X_1,X_2]=-X_2-X_3,
[X_3,X_1]=X_2+X_3$}&
 & {\scriptsize $dx_1[X_1-(x_2+x_3)(X_2+X_3)]+dx_2X_2+dx_3(X_3$}\\

\vspace{1mm}

& &  & {\scriptsize $+x_4 X_4)+dx_4X_4$}\\

\vspace{1mm}

{\scriptsize ${ 2A_2}$}& {\scriptsize$[X_1,X_2]=X_2,[X_3,X_4]=X_4$}&
 &{\scriptsize$dx_1(X_1+x_2X_2)+dx_2X_2+dx_3(X_3+x_4X_4)+dx_4X_4$}\\

\vspace{1mm}

& &  & {\scriptsize $$}\\

\vspace{1mm}

{\scriptsize ${ II+R}$}& {\scriptsize$[X_2,X_3]=X_1$}&
 &
{\scriptsize $dx_1X_1+dx_2(X_2+x_3X_1)+dx_3X_3+dx_4X_4$} \\

\vspace{-1mm}

{\scriptsize $IV+R$}&
{\scriptsize$[X_1,X_2]=-X_2+X_3,[X_1,X_3]=-X_3$}& & {\scriptsize $dx_1 [X_1-x_2X_2+(x_2-x_3)X_3]+dx_2X_2+dx_3X_3$}\\

\vspace{1mm}

& &  & {\scriptsize $+dx_4X_4$}\\

\vspace{1mm}

{\scriptsize $V+R$}&
{\scriptsize$[X_1,X_2]=-X_2,[X_1,X_3]=-X_3$}&  &
{\scriptsize $dx_1\left(X_1-x_2X_2-x_3X_3\right)+dx_2X_2+dx_3X_3+dx_4X_4$}\\

\vspace{-1mm}

{\scriptsize $VI_0+R$}&
{\scriptsize$[X_1,X_3]=X_2,[X_2,X_3]=X_1$}& &
{\scriptsize $dx_1(X_1\cosh{x_3}+X_2\sinh{x_3})+dx_2(X_2\cosh{x_3}$}\\

\vspace{1mm}

& &  & {\scriptsize$+X_1\sinh{x_3})+dx_3X_3+dx_4X_4$}\\

\vspace{-1mm}

{\scriptsize $VI_a+R$}&
{\scriptsize$[X_1,X_2]=-aX_2-X_3,[X_3,X_1]=X_2+aX_3$}& &
{\scriptsize $dx_1[X_1-(ax_2+x_3)X_2-(ax_3+x_2)X_3]+dx_2X_2$}\\

\vspace{1mm}

& &  & {\scriptsize $+dx_3X_3+dx_4X_4$}\\

\vspace{-1mm}

{\scriptsize $VII_0+R$}&
{\scriptsize$[X_1,X_3]=-X_2,[X_2,X_3]=X_1$}&  &
{\scriptsize $dx_1(X_1\cos{x_3}-X_2\sin{x_3})+dx_2(X_2\cos{x_3}+X_1\sin{x_3})$}\\

\vspace{1mm}

& &  & {\scriptsize$+dx_3X_3+dx_4X_4$}\\

\vspace{-1mm}

{\scriptsize $VII_a+R$}&
{\scriptsize$[X_1,X_2]=-aX_2+X_3,[X_3,X_1]=X_2+aX_3$}&  &
{\scriptsize $dx_1[X_1-(ax_2+x_3)X_2-(-ax_3+x_2)X_3]+dx_2X_2$}\\

\vspace{1mm}

& &  & {\scriptsize $+dx_3X_3+dx_4X_4$}\\

\vspace{-1mm}

{\scriptsize $VIII+R$}&
{\scriptsize$[X_1,X_3]=-X_2,[X_2,X_3]=X_1,[X_1,X_2]=-X_3$}&
 &
{\scriptsize $dx_1(X_1\cosh{x_2}\cos{x_3}-X_2\cosh{x_2}\sin{x_3}-X_3\sinh{x_2})$}\\

\vspace{1mm}

& &  &{\scriptsize$+dx_2(X_2\cos{x_3}+X_1\sin{x_3})+dx_3X_3+dx_4X_4$}\\

\vspace{-1mm}

{\scriptsize $IX+R$}&
{\scriptsize$[X_1,X_3]=-X_2,[X_2,X_3]=X_1,[X_1,X_2]=X_3$}&
 &
{\scriptsize $dx_1(X_1\cos{x_2}\cos{x_3}-X_2\cos{x_2}\sin{x_3}+X_3\sin{x_2})$}\\

\vspace{1mm}

& &  &{\scriptsize$+dx_2(X_2\cos{x_3}+X_1\sin{x_3})+dx_3X_3+dx_4X_4$}\\

\vspace{-1mm}

{\scriptsize${A_{4,1}}$}&{\scriptsize$[X_2,X_4]=X_1,[X_3,X_4]=X_2$}&
 &{\scriptsize $dx_1X_1+dx_2(X_1x_4+X_2)+dx_3(X_3+\frac{x_4^2}{2}X_1+X_2x_4)$}\\

\vspace{1mm}

& &  &{\scriptsize$+dx_4X_4$}\\

\vspace{1mm}

{\scriptsize${{A^a}_{4,2}}$}&{\scriptsize$[X_1,X_4]=aX_1,[X_2,X_4]=X_2,[X_3,X_4]=X_2+X_3$}&
 &
{\scriptsize $dx_1X_1e^{ax_4}+dx_2X_2e^{x_4}+dx_3(X_3+X_2)e^{x_4}+dx_4X_4$}\\

\vspace{-1mm}

{\scriptsize ${{A^1}_{4,2}}$}&
{\scriptsize$[X_1,X_4]=aX_1,[X_1,X_4]=X_1
,[X_2,X_4]=X_2,$}& &
{\scriptsize $dx_1X_1e^{x_4}+dx_2X_2e^{x_4}+dx_3(X_3+X_2)e^{x_4}+dx_4X_4$}\\

\vspace{1mm}

{\scriptsize
$$}&{\scriptsize$[X_3,X_4]=X_2+X_3$}&{\scriptsize$$}&{\scriptsize$$}\\

\vspace{1mm}

{\scriptsize ${A_{4,3}}$}&
{\scriptsize$[X_1,X_4]=X_1,[X_3,X_4]=X_2$}&&
{\scriptsize $dx_1X_1e^{x_4}+dx_2X_2+dx_3(X_3+x_4X_2)+dx_4X_4$}\\

\vspace{-1mm}

{\scriptsize ${A_{4,4}}$}&
{\scriptsize$[X_1,X_4]=X_1,[X_2,X_4]=X_1+X_2,$}&&
{\scriptsize$dx_1X_1e^{x_4}+dx_2(X_2+x_4X_1)e^{x_4}+dx_3(X_3+X_2x_4$}\\

\vspace{1mm}

{\scriptsize $$}&{\scriptsize$[X_3,X_4]=X_2+X_3$}&{\scriptsize$$}&{\scriptsize$+X_1x_4^2/2)e^{x_4}+dx_4X_4$}\\

\vspace{1mm}

{\scriptsize ${{A^{a,b}}_{4,5}}$}&
{\scriptsize$[X_1,X_4]=X_1,[X_2,X_4]=aX_2,[X_3,X_4]=bX_3$}&&
{\scriptsize $dx_1X_1e^{x_4}+dx_2X_2e^{ax_4}+dx_3X_3e^{bx_4}+dx_4X_4$}\\

\vspace{1mm}

{\scriptsize ${{A^{a,a}}_{4,5}}$}&
{\scriptsize$[X_1,X_4]=X_1,[X_2,X_4]=aX_2,[X_3,X_4]=aX_3$}&&
{\scriptsize $dx_1X_1e^{x_4}+dx_2X_2e^{ax_4}+dx_3X_3e^{ax_4}+dx_4X_4$}\\

\vspace{1mm}

{\scriptsize ${{A^{a,1}}_{4,5}}$}&
{\scriptsize$[X_1,X_4]=X_1,[X_2,X_4]=aX_2,[X_3,X_4]=X_3$}&&
{\scriptsize $dx_1X_1e^{x_4}+dx_2X_2e^{ax_4}+dx_3X_3e^{x_4}+dx_4X_4$}\\

\vspace{1mm}

{\scriptsize ${{A^{1,1}}_{4,5}}$}&
{\scriptsize$[X_1,X_4]=X_1,[X_2,X_4]=X_2,[X_3,X_4]=X_3$}&&
{\scriptsize $dx_1X_1e^{x_4}+dx_2X_2e^{x_4}+dx_3X_3e^{x_4}+dx_4X_4$}\\

\vspace{-1mm}

{\scriptsize ${{A^{a,b}}_{4,6}}$}&
{\scriptsize$[X_1,X_4]=aX_1,[X_2,X_4]=2X_2-X_3,$}&&

{\scriptsize $dx_1X_1e^{ax_4}+e^{bx_4}[dx_2(X_2\cos x_4-X_3\sin x_4)$}\\

\vspace{-1mm}

{\scriptsize $$}&{\scriptsize$[X_3,X_4]=X_2+bX_3$}&{\scriptsize
$$}&{\scriptsize$+dx_3(X_3\cos x_4+X_2\sin
x_4)]+dx_4X_4$}\\\smallskip \\

\hline\hline
    \end{tabular}
\end{center}
\newpage

\vspace{4mm}
\begin{center}
    \begin{tabular}{l l l l  l l p{25mm} }
    \hline\hline
   \vspace{-1mm}
{\scriptsize ${\bf Algebra}$ }& {\scriptsize Non-zero commutation
relations }&&{\scriptsize $g^{-1}
dg$}\\\hline

\vspace{-1mm}

{\scriptsize ${{A}_{4,7}}$}&
{\scriptsize$[X_1,X_4]=2X_1,[X_2,X_4]=X_2,$}&&

{\scriptsize $dx_1X_1e^{2x_4}+dx_2(X_2e^{x_4}-x_3X_1e^{2x_4})+dx_3e^{x_4}(x_4X_2$}\\

\vspace{1mm}

{\scriptsize$$}&{\scriptsize$[X_3,X_4]=X_2+X_3,[X_2,X_3]=X_1$}&{\scriptsize$$}&{\scriptsize$+X_3)+dx_4X_4$}\\

\vspace{1mm}

{\scriptsize ${{A}_{4,8}}$}&
{\scriptsize$[X_2,X_4]=X_2,[X_3,X_4]=-X_3,[X_2,X_3]=X_1$}&&
{\scriptsize $dx_1X_1+dx_2(X_2e^{x_4}+x_3X_1)+dx_3e^{-x_4}X_3+dx_4X_4$}\\

\vspace{-1mm}

{\scriptsize ${{A}^b_{4,9}}$}&
{\scriptsize$[X_1,X_4]=(1+b)X_1,[X_2,X_4]=X_2,$}&&
{\scriptsize $dx_1X_1e^{(1+b)x_4}+dx_2(X_2e^{x_4}+x_3e^{(1+b)x_4}X_1)$}\\

\vspace{1mm}

{\scriptsize
$$}&{\scriptsize$[X_3,X_4]=bX_3,[X_2,X_3]=X_1$}&{\scriptsize $$}&
{\scriptsize$+dx_3e^{x_4}(x_4X_2+X_3)+dx_4X_4$}\\

\vspace{-1mm}

{\scriptsize${{A}^1_{4,9}}$}&{\scriptsize$[X_1,X_4]=2X_1,[X_2,X_4]=X_2,[X_3,X_4]=X_3$},&&

{\scriptsize$dx_1X_1e^{2x_4}+dx_2(X_2e^{x_4}+x_3e^{2x_4}X_1)+dx_3e^{x_4}X_3+dx_4X_4$}\\

\vspace{1mm}

& {\scriptsize$[X_2,X_3]=X_1$}\\

\vspace{2mm}

{\scriptsize${{A^0}_{4,9}}$}&{\scriptsize$[X_1,X_4]=X_1,[X_2,X_4]=X_2,[X_2,X_3]=X_1$}&
& {\scriptsize $dx_1X_1e^{x_4}+dx_2(X_2e^{x_4}+x_3e^{x_4}X_1)+dx_3X_3+dx_4X_4$}\\

\vspace{-1mm}

{\scriptsize$A_{4,10}$}&{\scriptsize$[X_2,X_4]=-X_3,\;[X_3,X_4]=X_2,;[X_2,X_3]=X_1$}&
 & {\scriptsize $dx_1X_1+dx_2(x_3X_1+X_2\cos{x_4}-X_3\sin{x_4})+dx_3(X_3\cos{x_4}$}\\

\vspace{1mm}

& &  & {\scriptsize $+X_2\sin{x_4})+dx_4X_4$}\\

\vspace{-1mm}

{\scriptsize${{A^a}_{4,11}}$}&{\scriptsize$[X_1,X_4]=2aX_1,[X_2,X_4]=aX_2-X_3,$}&&

{\scriptsize $dx_1X_1e^{2ax_4}+dx_2(x_3e^{2ax_4}X_1+X_2e^{ax_4}\cos{x_4}-$}\\

\vspace{1mm}

{\scriptsize$$}&{\scriptsize$[X_3,X_4]=X_2+aX_3,[X_2,X_3]=X_1$}&{\scriptsize$$}&{\scriptsize$X_3e^{ax_4}\sin{x_4})+dx_3(X_3e^{ax_4}\cos{x_4}+X_2e^{ax_4}\sin{x_4})+dx_4X_4$}\\

\vspace{-1mm}

{\scriptsize ${A_{4,12}}$}&
{\scriptsize$[X_1,X_4]=-X_2,[X_1,X_3]=X_1,[X_2,X_4]=X_1,$}&&

{\scriptsize $dx_1e^{x_3}(X_1\cos{x_4}-X_2\sin{x_4})+dx_2e^{x_3}(X_2\cos{x_4}+X_1\sin{x_4})$}\\

\vspace{-1mm}

& {\scriptsize$[X_2,X_3]=X_2$}&{\scriptsize$$}&{\scriptsize$+dx_3X_3+dx_4X_4$}\smallskip \\

\hline\hline
\end{tabular}
\end{center}
\newpage
\vspace{-60mm}{\bf AppendixB:}

Here the results of section 5 of ref \cite{Mojaveri} are corrected as follows:
By using (13) and (15) and (54)of ref \cite{Mojaveri} we have:
\renewcommand\theequation{\arabic{tempeq}\alph{equation}}
\setcounter{equation}{-1} \addtocounter{tempeq}{1}
\begin{eqnarray}
\tau=\frac{1}{\gamma}\ln(\frac{\gamma t}{\delta}),
\end{eqnarray}
with $\gamma=\frac{p_1+p_3+p_4-2N}{2}$ and $\delta=(L_1L_2^2L_3^2)^\frac{1}{4}$. Then using relation (20) we have:
\begin{equation}
k^2 {T_{00}}^\phi=\frac{N^2}{3\gamma^2 t^2}(1-\frac{1}{2}(\frac{\gamma t}{\delta})^\frac{2N}{3\gamma})
\end{equation}
\begin{equation}
k^2 {T_{11}}^\phi=k^2 {T_{22}}^\phi=\frac{\sqrt{L_1}N^2}{6\gamma^2 t^2}(\frac{\gamma t}{\delta})^\frac{3p_1-2N}{6\gamma}
\end{equation}
\begin{equation}
k^2 {T_{33}}^\phi=\frac{L_2 N^2}{6\gamma^2 t^2}(\frac{\gamma t}{\delta})^\frac{3p_3-N}{3\gamma}
\end{equation}
\begin{equation}
k^2 {T_{44}}^\phi=\frac{L_3 N^2}{6\gamma^2 t^2}(\frac{\gamma t}{\delta})^\frac{3p_4-N}{3\gamma}
\end{equation}
After comparison of the above results with (57) of ref \cite{Mojaveri} we must have $\frac{p_1}{2}=p_3=p_4$ and $\sqrt{L_1}=L_2=L_3$ (i.e $a_1=a_2=a_3=a_4$)
then from conditions (55) of ref \cite{Mojaveri} we have $p_1=\frac{4}{3}N$ so $\gamma=\frac{N}{3}$. Note that from (26) we must have $\gamma>0$ so $N>0$.
Now using equation of state as $P(t)=\omega(t) \rho(t)$ we have the following results:
\begin{equation}
\rho(t)=\frac{3}{2t^2{L_1}^{\frac{3}{2}}}(-2{L_1}^{\frac{3}{2}}+ \frac{N^2 t^2}{9}),
\end{equation}
\begin{equation}
\omega(t)=\frac{{L_1}^{\frac{5}{4}}Nt}{3}(-2{L_1}^{\frac{3}{2}}+ \frac{N^2 t^2}{9})^{-1},
\end{equation}
\begin{equation}
P(t)=\frac{N}{2{L_1}^{\frac{1}{4}}t},
\end{equation}
\begin{equation}
a_i(t)=\frac{N^2 t^2}{9L_1},
\end{equation}
and for deceleration parameter we have $q_i(t)=-\frac{1}{2}$.

\newpage
\subsection*{Acknowledgements}

This work has been supported by the research vice chancellor of Azarbaijan Shahid Madani University under research fund No. 1402/231.

\end{document}